%% file: main.tex
\documentclass{article}

    \usepackage[preprint]{neurips_2025}

\usepackage[utf8]{inputenc} 
\usepackage[T1]{fontenc}    
\usepackage{hyperref}       
\usepackage{url}            
\usepackage{booktabs}       
\usepackage{amsfonts}       
\usepackage{nicefrac}       
\usepackage{microtype}      
\usepackage{xcolor}         
\usepackage{amssymb,mathrsfs,amsmath}
\usepackage{amsmath}
\usepackage{subcaption}
\usepackage{float}
\usepackage{tablefootnote}
\usepackage{balance}

\usepackage{graphicx}
\usepackage{enumitem}
\usepackage{multirow}
\usepackage{wrapfig} 
\usepackage[table]{xcolor}  

\definecolor{lightgreen}{RGB}{220, 255, 220}
\definecolor{rowgray}{gray}{0.95}

\title{$i$MIND: Insightful Multi-subject Invariant Neural Decoding}

\author{%
  Zixiang Yin, Jiarui Li, $^*$Zhengming Ding \\
  Department of Computer Science, Tulane University \\
  \texttt{\{zyin, jli78, zding1\}@tulane.edu} \\
  $^*$Corresponding author.\\
}

\begin{document}

\maketitle

\input{sec/M0_abstract}    

\input{sec/M1_intro}
\input{sec/M3_method}

\input{sec/M4_results}
\input{sec/M5_discussion}

\bibliographystyle{plain}
\bibliography{main}


\newpage
\appendix
\input{sec/A1_appendix}


\newpage
\input{sec/A2_checklist}

\end{document}

%% file: sec/M0_abstract.tex
\begin{abstract}
Decoding visual signals holds the tantalizing potential to unravel the complexities of cognition and perception. While recent studies have focused on reconstructing visual stimuli from neural recordings to bridge brain activity with visual imagery, existing methods offer limited insights into the underlying mechanisms of visual processing in the brain. To mitigate this gap, we present an \textit{i}nsightful \textbf{M}ulti-subject \textbf{I}nvariant \textbf{N}eural \textbf{D}ecoding ($i$MIND) model, which employs a novel dual-decoding framework--both biometric and semantic decoding--to offer neural interpretability in a data-driven manner and deepen our understanding of brain-based visual functionalities. Our $i$MIND model operates through three core steps: establishing a shared neural representation space across subjects using a ViT-based masked autoencoder, disentangling neural features into complementary subject-specific and object-specific components, and performing dual decoding to support both biometric and semantic classification tasks. Experimental results demonstrate that $i$MIND achieves state-of-the-art decoding performance with minimal scalability limitations. Furthermore, $i$MIND empirically generates voxel-object activation fingerprints that reveal object-specific neural patterns and enable investigation of subject-specific variations in attention to identical stimuli. These findings provide a foundation for more interpretable and generalizable subject-invariant neural decoding, advancing our understanding of the voxel semantic selectivity as well as the neural vision processing dynamics.
\end{abstract}

%% file: sec/M1_intro.tex
\section{Introduction}
\label{sec:intro}

Deep learning models have achieved remarkable success across a wide range of computer vision tasks \cite{cv1,cv2,cv3,cv4,cv5,cv6}, driven by their powerful feature extraction capabilities and their ability to learn complex patterns from large-scale datasets. In these contexts, annotated visual labels often serve as a form of crowdsourced semantic knowledge \cite{crowdsourcing}. However, in many real-world applications—such as autonomous driving \cite{auto-drive1, auto-drive2, auto-drive3} and medical diagnosis \cite{disease1, disease2}—annotations tend to be noisy, inconsistent, or ambiguous due to the inherent subjectivity of human interpretation. To address these limitations, brain signals—particularly functional Magnetic Resonance Imaging (fMRI)—have been investigated as an alternative supervisory signal \cite{grill2004human}, offering the potential to model individual perceptual \cite{baldassarre2012individual}, emotional \cite{phan2002functional}, and attentional \cite{mangun1998erp} responses to visual stimuli.

Most recent research has focused on reconstructing visual stimuli from brain activity by projecting neural representations into deep visual spaces and employing generative models such as GANs or diffusion models \cite{data_prep2, mind-bridge, mindeye1, mindreader, mind-vis}. These approaches have yielded visually plausible, high-resolution images, suggesting that deep learning can serve as a bridge between brain activity and imagery. \textbf{\textit{However, despite these aesthetic achievements, we argue that reconstruction alone is fundamentally inadequate for understanding brain vision mechanism.}} Specifically, the following limitations exist:
\begin{itemize}[leftmargin=*]
    \item \textbf{Reconstruction relies heavily on pretrained generative priors}, which often dominate the decoding process and may introduce model-specific biases that obscure genuine neural content.\vspace{-1mm}
    \item \textbf{Brain recording  may not encode all fine-grained details necessary for accurate reconstruction}—especially across subjects—rendering pixel-level reconstruction ill-posed and often misleading. This ultimately shifts the focus from reconstruction to generation.\vspace{-1mm}
\end{itemize}

Thus, we argue that reconstructing visual stimuli is neither a reliable nor interpretable strategy for decoding neural representations. A more effective alternative is to classify the subject's perceptual experience directly from fMRI data, enabling the identification of visual concepts embedded in neural responses \cite{clip_mused}. This discriminative classification-based approach supports evaluation through standard metrics and allows researchers to disentangle both shared and individual-specific semantic components of brain activity—capabilities that reconstruction methods often fail to provide. 

Technically, current neural decoding solutions follow two main strategies: single-subject models \cite{brain-scuba, mind-vis, data_prep2, mindreader, mindeye1, brain-dive}, which suffer from data scarcity, overfitting, and poor scalability due to the high cost of fMRI collection; and multi-subject models \cite{clip_mused, mind-bridge, emb, mindeye2}\footnote{See Appendix \ref{subsec:realte_works} for more related work.}, which face substantial inter-subject variability from anatomical and functional differences. This can lead to entangled neural representations, where subject-specific and object-related signals become mixed \cite{dubois2016building}, degrading decoding performance and interpretability. This raises a key research question: \vspace{-2mm}
\begin{quote}
\textit{``How can we develop a discriminative neural decoding framework that generalizes across individuals while preserving subject-specific semantic interpretations of visual stimuli?''}\vspace{-2mm}
\end{quote}

To answer this, we present a novel approach to decoding brain activity using deep classification models, aiming to capture shared neural representations across individuals while preserving personalized interpretations shaped by diverse experiences and backgrounds. Toward this goal, we introduce the \textbf{i}nsightful \textbf{M}ulti-subject \textbf{I}nvariant \textbf{N}eural \textbf{D}ecoding (\textbf{$i$MIND}) model—a dual-purpose framework that supports both \textit{biometric decoding} (identifying individuals) and \textit{semantic decoding} (classifying perceived objects). By constructing voxel-object activation maps and examining subject-specific attentional patterns in response to complex, multi-object scenes, $i$MIND offers a principled path to disentangling individual and shared components of visual perception, ultimately advancing our understanding of brain-vision mechanisms. To sum up, our contributions are threefold:\vspace{-2mm}
\begin{itemize}[leftmargin=*]
\item We introduce a multi-subject dual-decoding framework that disentangles neural signals into subject-specific and object-specific components, enabling scalable and subject-aware brain analysis with state-of-the-art semantic and biometric decoding performance.
\item We develop a visual-neural interaction module that identifies object-voxel activation patterns in a data-driven manner, revealing how different subjects encode object-level semantics in the brain.
\item We perform a comprehensive analysis of attention-based variability across subjects viewing the same visual input, providing insights into individualized neural responses under time-constrained conditions.
\end{itemize}

%% file: sec/M3_method.tex
\section{Proposed Method}
\label{sec:method}

\subsection{Problem Formulation}

Consider a neural dataset containing brain activities of subjects $\mathcal{S}$ in response to visual stimuli drawn from an image dataset with $M$ samples $\mathcal{D}=\{(\mathbf{X}_i, \mathbf{y}_i)\}_{i=1}^{M}$, where each visual stimulus $\mathbf{X}_i$ is a three-channel image with a resolution of $H\times W$, paired with a \textbf{non-exclusive} ground-truth label $\mathbf{y}_i\in\mathbb{R}^C$ representing $C$ object categories. During neural recordings, all images $\mathbf{X}$ are viewed by a group of subjects $\mathcal{S}$.
For each subject $s\in \mathcal{S}$ viewing the $i$-th image $\mathbf{X}_i$, an fMRI voxel response $\mathbf{V}_{i,s}$ is recorded, capturing neural activity specific to the subject-image pair. Note that the voxel signal $\mathbf{V}_{i,s}$ is flattened as an 1D vector with various lengths (12,682 $\sim$ 17,907) across subjects due to biometric differences in neural structure. Following \citep{mind-vis}, we apply wrap-around padding to achieve a uniform voxel length $L$. Details on pre-processing procedures are provided in Appendix \ref{subsec:preprocess}.

\subsection{Framework Overview}

Our goal is to decouple arbitrary fMRI voxel signals into subject-invariant and subject-specific components from both semantic and biometric perspectives by developing a visual-neural model $\mathcal{M}$. Formally, this can be expressed as a mapping:
\begin{align}
    \mathcal{M}: \underbrace{\mathbb{R}^{H \times W \times 3}}_{\text{image}} \times \underbrace{\mathbb{R}^L}_{\text{voxels}} \rightarrow \underbrace{\mathbb{R}^C}_{\text{semantics}} \times \underbrace{\mathbb{R}^{|\mathcal{S}|}}_{\text{biometrics}},
\end{align}
where \textit{semantic decoding} seeks to extract object-related neural representations that are consistent across subjects, whereas \textit{biometric decoding} focuses on capturing subject-specific neural patterns that are independent of the visual stimuli.

To address both tasks, we propose the \textbf{i}nsightful \textbf{M}ulti-subject \textbf{I}nvariant \textbf{N}eural \textbf{D}ecoding model, abbreviated as \textbf{\textit{i}MIND}. Building on the SC-MBM framework~\citep{mind-vis}, our method constructs a $d$-dimensional shared latent neural space $\mathcal{F}$ across subjects to reduce the noise and redundancy inherent in fMRI signals~\citep{fmri_noise1}. Specifically, we employ a Vision Transformer (ViT)-based encoder $\mathcal{E} : \mathbb{R}^L \rightarrow \mathbb{R}^{N \times d}$ that projects each input voxel signal $\mathbf{V}$ into a set of $N$ neural feature tokens, denoted as $\mathbf{F} = \{\mathbf{f}_j\}_{j=1}^N$, where each $\mathbf{f}_j$ is a $d$-dimensional \textit{neural feature token}. This yields $N$ neural representations per fMRI input, all embedded in the shared latent space $\mathcal{F}$. The encoder $\mathcal{E}(\cdot)$ is pretrained using a masked autoencoding objective in a self-supervised manner, emphasizing voxel-wise reconstruction.

Subsequently, we disentangle the learned features into object-specific and subject-specific components. The subject-specific features are used for subject classification (biometric decoding), while the object-specific features are aligned with frozen CLIP visual embeddings of the corresponding images via a cross-attention mechanism, enabling multi-label object classification (semantic decoding).

\subsection{Subject-Object Disentanglement}\label{subsec:method_so_dis}


The self-supervised pretraining reduces noise and redundancy in fMRI signals by encouraging the model to capture generalizable patterns. However, the resulting neural features $\mathcal{F}$ are not explicitly tailored for downstream decoding tasks such as biometric identification or semantic classification. Because the reconstruction objective treats all latent information as equally relevant, it often leads to entangled representations—mixing subject-specific and object-specific components. This entanglement limits the interpretability of the learned features and hinders task-specific performance, especially when distinguishing between individual variability and shared visual semantics is essential.

Mathematically, this can be formalized by interpreting each neural feature token $\mathbf{f} = (f_1, f_2, \dots, f_d) \in \mathbb{R}^d$ as the \textbf{coordinate representation} of a corresponding neural point $\mathbf{p}$ in the latent neural feature space $\mathcal{F}$. By default, this representation is expressed with respect to the standard basis $\mathbf{E} = \{\mathbf{e}_k \in \mathbb{R}^d\}|_{k=1}^d$ of $\mathbb{R}^d$, denoted as $[\mathbf{p}]_{\mathbf{E}}$:\vspace{-2mm}
\begin{align}
[\mathbf{p}]_{\mathbf{E}} := \mathbf{f} = \sum_{k=1}^d f_k \mathbf{e}_k \in \mathbb{R}^d.
\end{align}

From this perspective, the entanglement of subject-specific and object-specific information within $\mathbf{f}$ can be attributed to the default use of $\mathbf{E}$ as the basis for spanning the neural feature space $\mathcal{F}$. As $\mathbf{E}$ is implicitly determined by the self-supervised task in the first training stage, this choice is beyond direct control, resulting in an inevitable blending of both subject and object information within the feature representation $\mathbf{f}$ for any neural point $\mathbf{p}\in\mathcal{F}$.

To enable effective downstream biometric and semantic decoding, we propose a solution from the perspective of feature disentanglement \cite{trauble2021disentangled,dittadi2021transfer,fumero2023leveraging,zhang2024identifiable}. We begin with assuming that the subject-specific and object-specific information within a neural point $\mathbf{p}$ are \textbf{linearly} entangled in the current neural feature representation $\mathbf{f}$ under the basis $\mathbf{E}$. While this may not fully characterize the complexities of neural dynamics, it serves as a simplified approximation to provides a meaningful step toward understanding the interplay between subject-specific and object-specific information. More crucially, this assumption applies at the \textbf{latent feature level}, not at the original fMRI signal level. At this level, the assumption is reasonable, as it aligns with the principles of deep classification tasks, where linear MLP classifiers rely on deep neural networks to transform inputs into linearly separable features for accurate classification. Based on this assumption, we resolve the linear entanglement by re-representing $\mathbf{p}$ with respect to a new basis $\mathbf{B}$. Specifically, we seek a \textbf{learnable} basis $\mathbf{B}=(\mathbf{B}_\text{subj},\mathbf{B}_\text{obj})$ of $\mathbb{R}^d$ that perfectly separates the subject-specific and object-specific features. Mathematically, this re-representation would allow $\mathbf{z}$, the representation of the same neural point $\mathbf{p}$ with respect to the new basis $\mathbf{B}$, to be distinctly split into subject-specific $\mathbf{z}_\text{subj}$ and object-specific $\mathbf{z}_\text{obj}$ components:
\begin{align}\label{eq:sep}
    \mathbf{z} &= [\mathbf{p}]_\mathbf{B} =\bigl(\left[\mathbf{p}\right]_{\mathbf{B}_\text{subj}}, \left[\mathbf{p}\right]_{\mathbf{B}_\text{obj}}\bigr) =(\mathbf{z}_\text{subj}, \mathbf{z}_\text{obj})\in\mathbb{R}^d.
\end{align}
 
According to the mathematical property of bases, the separation of $\mathbf{z}_\text{subj}$ and $ \mathbf{z}_\text{obj}$ is guaranteed as long as $\mathbf{B}$ forms an \textbf{orthonormal} basis of $\mathbb{R}^d$, i.e., $\mathbf{B}\mathbf{B}^\top = \mathbf{I}_d$, where $\mathbf{I}_d$ is the identity matrix of rank $d$. With this orthonormality condition satisfied, the transformation from the original representation of $\mathbf{f}$ to the new representation $\mathbf{z}$ of the same neural point $\mathbf{p}$ can be derived as follows:
\begin{align}
    \mathbf{z} = [\mathbf{p}]_\mathbf{B} = \mathbf{B}^{-1}[\mathbf{p}]_\mathbf{E} = \mathbf{B}^\top[\mathbf{p}]_\mathbf{E} = \mathbf{B}^\top\mathbf{f}\in\mathbb{R}^d.
\end{align}
Combined with Eq. \eqref{eq:sep}, we finally arrive at our subject-object disentanglement formulation:
\begin{align}\label{eq:token_sep}
\mathbf{z}_\text{subj}=\mathbf{B}^\top_\text{subj}\mathbf{f}\in\mathbb{R}^{d_\text{subj}}\quad\text{and}\quad\mathbf{z}_\text{obj} = \mathbf{B}^\top_\text{obj}\mathbf{f}\in\mathbb{R}^{d_\text{obj}}.
\end{align}

From a feature transformation perspective, we realize subject-object disentanglement by decomposing the original neural space $\mathcal{F}$ into two complementary (orthogonal) subspaces: the subject-specific neural subspace $\mathcal{F}_\text{subj}$ and the object-specific neural subspace $\mathcal{F}_\text{obj}$. This disentanglement is achieved by learning two orthonormal sets, $\mathbf{B}_\text{subj}\in\mathbb{R}^{d\times d_\text{subj}}$ and $\mathbf{B}_\text{obj}\in\mathbb{R}^{d\times d_\text{obj}}$, which span $\mathcal{F}_\text{subj}$ and $\mathcal{F}_\text{obj}$ respectively. In our framework, $d_\text{obj}$ is treated as a user-defined value, with $d_\text{subj}:=d-d_\text{obj}$ to complete the basis. A formal and theoretic proof is provided in Appendix \ref{subsec:bases_proof}.

The complementary relationship, $\mathcal{F}=\mathcal{F}_\text{subj}\oplus\mathcal{F}_\text{obj}$, guarantees a clear separation of subject and object information in the transformed neural representation $\mathbf{z}=(\mathbf{z}_\text{subj},\mathbf{z}_\text{obj})\in\mathbb{R}^d$ for each neural point $\mathbf{p}\in\mathcal{F}$, establishing a foundation for the subsequent biometric and semantic decoding tasks.

\subsection{Biometric \& Semantic Decoding}
In this section, we describe our approach to decoding fMRI signals biometricly and semantically.Note that an fMRI signal $\mathbf{V}$ is represented by $\mathbf{F}$ as a set of $N$ neural tokens $\{\mathbf{f}_j\}_{j=1}^N$ within the latent neural space $\mathcal{F}$. Based on Eq. \eqref{eq:token_sep}, the subject-specific feature map $\mathbf{Z}_\text{subj}\in\mathbb{R}^{N\times 
d_\text{subj}}$ and object-specific feature map $\mathbf{Z}_\text{obj}\in\mathbb{R}^{N\times 
d_\text{obj}}$ are generated from $\mathbf{F}\in\mathbb{R}^{N\times d}$ as follows:
\begin{align}\label{eq:fmap_sep}
\mathbf{Z}_\text{subj}=\mathbf{FB}_\text{subj}\quad\text{and}\quad\mathbf{Z}_\text{obj}=\mathbf{FB}_\text{obj}.
\end{align}

\noindent\textbf{Biometric Decoding}.
The biometric neural decoding is driven by a supervised multi-subject classification task. We apply a Global Average Pooling (GAP) operator, $\mathcal{G}_\text{subj}:\mathbb{R}^{N\times d_\text{subj}}\rightarrow\mathbb{R}^{d_\text{subj}}$, to the subject-specific neural feature map $\mathbf{Z}_\text{subj}$ to build a subject class token:
\begin{align}\label{eq:subj_cls}
    \mathbf{z}^\text{cls}_\text{subj}=\mathcal{G}_\text{subj}(\mathbf{Z}_\text{subj}).
\end{align}

Finally, a linear multi-subject classifier $\mathcal{C}_\text{subj}:\mathbb{R}^{d_\text{subj}}\rightarrow\mathbb{R}^S$ is applied for the final biometric prediction:
\begin{align}\label{eq:subj_pred}
\mathbf{\hat{y}}_\text{subj}=\mathcal{C}_\text{subj}(\mathbf{z}^\text{cls}_\text{subj}).
\end{align}

\noindent\textbf{Semantic Decoding}. To establish a feature-level connection between the fMRI voxel signals $\mathbf{V}$ and the semantic content of visual stimulus $\mathbf{X}$, we utilize the vision feature map $\mathbf{F_x}\in\mathbb{R}^{N_\mathbf{x}\times d_\mathbf{x}}$ from the last layer of a frozen CLIP visual encoder \cite{clip}. In contrast to most neural vision approaches that project fMRI features into the CLIP visual space \cite{data_prep2, mind-vis}, our method takes the opposite: treating the CLIP vision feature as a query to extract corresponding neural object features from $\mathbf{Z}_\text{obj}$. This design choice leverages the subject-invariant nature of the CLIP features, which, if properly pre-trained, can serve as a pseudo-ground-truth for objects' existence. This pseudo-ground-truth functions as an anchor for investigating variations in subjects' neural responses to different objects within the same visual stimulus. Technically, a multi-head cross-attention extractor $\mathcal{A}:\mathbb{R}^{N_\mathbf{x}\times d_\mathbf{x}}\times\mathbb{R}^{N\times d_\text{obj}}\times\mathbb{R}^{N\times d_\text{obj}}\rightarrow\mathbb{R}^{N_\mathbf{x}\times d_\text{obj}}$ is used and takes $\mathbf{Z}_\text{obj}$ in Eq. \eqref{eq:fmap_sep} as both key and value:
\begin{align}\label{eq:cross_attn}
\mathbf{Z}^\mathbf{F_x}_\text{obj}=\mathcal{A}\big(\text{Query}=\mathbf{F_x},\text{Key}=\mathbf{Z}_\text{obj},\text{Value}=\mathbf{Z}_\text{obj}\big).
\end{align}

In this way, we can implement our semantic decoding primarily on the fMRI modality, with the CLIP visual feature assisting in refining object-specific neural features for final multi-label object classification. Similar to what we have done in the biometric decoding in the previous section, a global feature operator $\mathcal{G}_\text{obj}:\mathbb{R}^{N_\mathbf{x}\times d_\text{obj}}\rightarrow\mathbb{R}^{d_\text{obj}}$ transforms $\mathbf{Z}^\mathbf{F_x}_\text{obj}$ into an object class token:
\begin{align}\label{eq:covecls}
    \mathbf{z}^\text{cls}_\text{obj}=\mathcal{G}_\text{obj}(\mathbf{Z}^\mathbf{F_X}_\text{obj}).
\end{align}
Following that a multi-label object classifier $\mathcal{C}_\text{obj}:\mathbb{R}^{d_\text{obj}}\rightarrow\mathbb{R}^C$ is applied for final semantic prediction:
\begin{align}\label{eq:obj_pred}
    \mathbf{\hat{y}}_\text{obj}=\mathcal{C}_\text{obj}(\mathbf{z}^\text{cls}_\text{obj}).
\end{align}

\subsection{Model Training}

The training process of our model consists of two stages. In the first stage, we follow the approach of SC-MBM \citep{mind-vis} to pre-train a ViT-based masked autoencoder for fMRI data, constructing a latent neural space via minimizing reconstruction error with a Mean-Square Error (MSE) loss. 

In the second stage, we retain only the fMRI encoder $\mathcal{E}(\cdot)$ from the first stage and optimize it with all other parameters in the proposed architecture. Three loss functions guide this stage. First, for biometric decoding, we introduce a subject classification loss $\mathcal{L}_\text{subj}$, which computes the cross-entropy $\mathcal{H}_\text{CE}$ with $\mathsf{softmax}$ activation against the one-hot label $\mathbf{y}_\text{subj}$ from the ground-truth subject index $s$: 
\begin{align}
\mathcal{L}_\text{subj}&:=\mathcal{H}_\text{CE}\big(\mathsf{softmax}\left(\mathbf{\hat{y}}_\text{subj}\right), \mathbf{y}_\text{subj}\big).
\end{align}

For multi-label semantic decoding, we employ an object loss $\mathcal{L}_\text{obj}$ which is a binary cross-entropy function $\mathcal{H}_\text{BCE}$ with sigmoid activation $\sigma(\cdot)$:
\begin{align}
\mathcal{L}_\text{obj}&:=\mathcal{H}_\text{BCE}\big(\sigma\left(\mathbf{\hat{y}}_\text{obj}\right), \mathbf{y}_\text{obj}\big).
\end{align}

Finally, we impose an orthonormal constraint on the learnable basis concatenation $\mathbf{B}=(\mathbf{B}_\text{subj},\mathbf{B}_\text{obj})$, as defined in Eq. \eqref{eq:fmap_sep}. This orthonormal loss $\mathcal{L}_\text{orth}$ ensures the perfect separation of subject-specific and object-subject features by minimizing $\mathcal{L}_\text{orth}:=\|\mathbf{BB}^\top-\mathbf{I}_d\|_\mathbf{F}^2$,
where $\left\Vert\cdot\right\Vert_\mathbf{F}$ indicates the Frobenius matrix norm, and $\mathbf{I}_d \in \mathbb{R}^{d\times{d}}$ is the identity matrix. 

In summary, in the second training stage, the total objective $\mathcal{L}$ is formulated as:
\begin{align}
\mathcal{L}=\mathcal{L}_\text{subj}+\mathcal{L}_\text{obj}+\lambda\mathcal{L}_\text{orth},
\end{align}
where $\lambda$ serves as a trade-off hyperparameter to balance the orthonormal constraint against the subject and object classification losses, which are considered equally important and share the same scale.

%% file: sec/M4_results.tex
\section{Experiments}
\label{sec:results}

\subsection{Experimental Setup}

\noindent\textbf{Dataset}. We evaluate our $i$MIND framework using the Natural Scenes Dataset (NSD) \citep{nsd}, a comprehensive, publicly available fMRI dataset capturing brain responses from 8 human subjects viewing natural scenes from MS-COCO \citep{mscoco}. Each subject passively viewed a set of 10,000 images for 3s, each repeated three times; 1,000 of these images were shared across all subjects, while the remaining 9,000 were unique to each individual, with no overlap between subjects. Due to incomplete sessions and data availability restrictions, not all trials are accessible for every subject, resulting in a total of 213,000 trials across all participants before pre-processing. In line with previous NSD studies \citep{data_prep1, data_prep2, mindeye1, mindeye2}, we used standardized train/test splits and averaged fMRI activations over repetitions for each image within each subject. This pre-processing yielded 69,566 training samples and 7,674 test samples, allowing us to train a single multi-subject model across all 8 subjects. Additional details on NSD, data pre-processing, and $i$MIND implementation can be found in Appendix \ref{subsec:implementation}. 


\begin{table}[ht]
\centering
\caption{Semantic decoding performance on the NSD dataset.}\label{table_sd}
\begin{tabular}{c|c|cccc}
\toprule
\rowcolor{rowgray} \textbf{Model Type} & \textbf{Methods} & \textbf{Modalities} & \textbf{mAP} $\uparrow$ & \textbf{AUC} $\uparrow$ & \textbf{Hamming} $\downarrow$ \\ \midrule
\multirow{2}{*}{Single-subject} & MLP \citep{clip_mused}*                & fMRI            & .258 & .854 & .033 \\
                                & ViT \citep{clip_mused}*                & fMRI            & .238 & .815 & .032 \\
\midrule
\multirow{6}{*}{Multi-subject}  & EMB \citep{emb}                       & fMRI            & .220 & .825 & .035 \\
                                & SMODEL-MLP \citep{clip_mused}*         & fMRI            & .150 & .767 & .039 \\
                                & SMODEL-ViT \citep{clip_mused}*         & fMRI            & .156 & .755 & .038 \\
                                & CLIP-MUSED \citep{clip_mused}         & fMRI+Image+Text & .258 & .877 & .030 \\ \cline{2-6}
                                & \multirow{2}{*}
                                {\textbf{$i$MIND (Ours)}} & fMRI            & \textbf{.310} & \textbf{.913} & \textbf{.027} \\ 
                                &                                       & fMRI+Image      & \textbf{.784} & \textbf{.984} & \textbf{.012} \\
\bottomrule
\multicolumn{6}{l}{\small * directly sourced from \citep{clip_mused} as benchmarks due to the limited research on semantic neural decoding} \\
\end{tabular} \vspace{-5mm}
\end{table}

\begin{figure}[ht]
  \centering
  \includegraphics[width=1\textwidth]{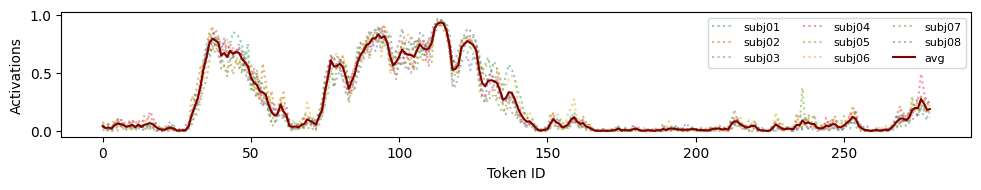}
  \vspace{-8mm}\caption{Average activations of tokens in $\mathbf{Z}_\text{obj}$ for the object \textit{Chair}.}\label{fig:tokens} \vspace{-6mm}
\end{figure}

\subsection{Neural Decoding Performance}

\noindent\textbf{Semantic Decoding}. For the semantic decoding task, we evaluate and compare with other models using three standard metrics for multi-label classification: mean Average Precision (mAP), the area under the receiver operating characteristic curve (AUC), and Hamming distance. Table \ref{table_sd} categorizes models based on their ability to process multi-subject fMRI signals simultaneously or on a per-subject basis, as well as the modalities used for object classification. While our $i$MIND model is designed to process both image and fMRI modalities, it can also be adapted as a single-modality model by simply removing the cross-attention mechanism in Eq. \eqref{eq:cross_attn} and using $\mathbf{Z}_\text{obj}$ in Eq. \eqref{eq:fmap_sep} directly for semantic decoding. Experimental results indicate that $i$MIND achieves superior performance across all three metrics, establishing a new state-of-the-art for semantic decoding in both single-modality and multi-modality settings.

\begin{wraptable}{r}{0.5\textwidth}
\centering\vspace{-4mm}
\caption{Biometric decoding performance on NSD.}
\label{table_bd}
\begin{tabular}{c|ccc}
    \toprule
    \rowcolor{rowgray} \textbf{Methods}           & \textbf{Distance/Loss} & \textbf{ACC} & \textbf{MCC}    \\ 
    \midrule
    \multirow{2}{*}{K-Means}   & Euclidean           & .181          & .068  \\
                               & Cosine              & .232          & .126  \\
    \midrule
    \multirow{2}{*}{MLP}       & MSE                 & .167          & .048  \\
                               & Cross Entropy       & .283          & .181  \\
    \midrule
    \rowcolor{lightgreen}
    \textbf{$i$MIND}           & Cross Entropy       & \textbf{.999} & \textbf{.999}  \\
    \bottomrule
\end{tabular}
\end{wraptable}


\noindent\textbf{Biometric Decoding}. To the best of our knowledge, no existing models support subject classification using fMRI. To provide a comprehensive evaluation, we established baseline models using the exact preprocessing steps from our proposed method and conducted both supervised and unsupervised biometric decoding, as presented in Table \ref{table_bd}. Top-1 accuracy (ACC) and Matthews Correlation Coefficient (MCC) \citep{mcc} are used as metrics. The poor performance of naive subject classification methods underscore the complexity of neural data across subjects, whereas the near-perfect classification achieved by $i$MIND highlights the effectiveness and necessity of our subject-object disentanglement approach. This demonstrates that within our framework, biometric fMRI feature are highly discriminative across subjects and our linearity assumption is reasonable. By facilitating the extraction of task-relevant features, our disentanglement method further enhances downstream semantic and biometric decoding. Details on how we build biometric decoding baselines are provided in Appendix \ref{subsec:subj_baseline}. 

\begin{figure}[t]
  \centering
  \includegraphics[width=1\textwidth]{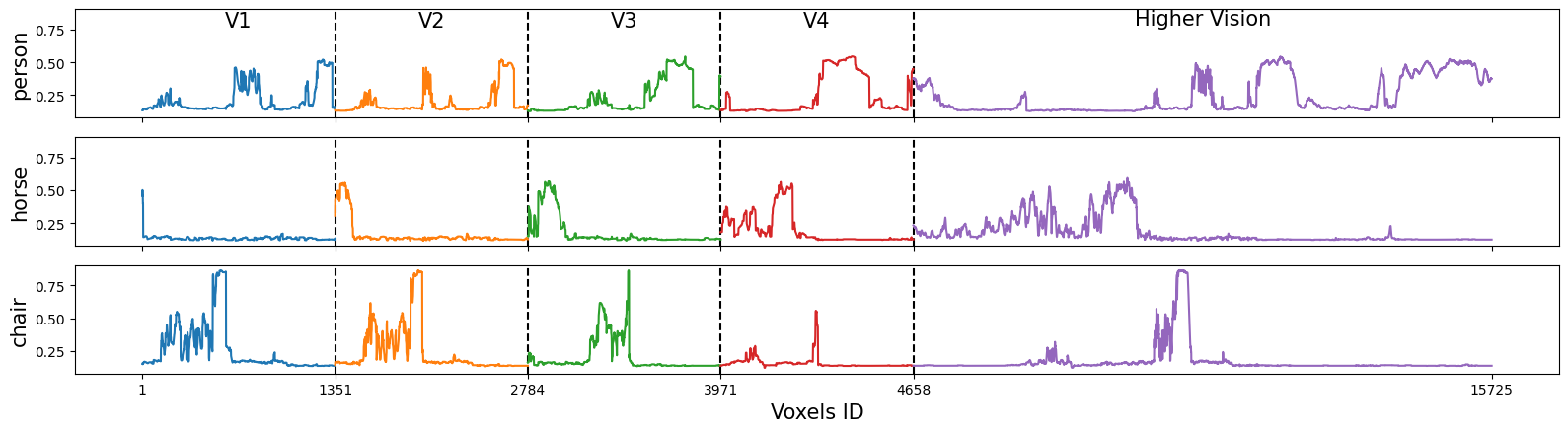}
  \vspace{-7mm}\caption{1D Object-Voxel activations by brain vision ROIs for subj01.}\vspace{-3mm}\label{fig:vo_median}
\end{figure}

\subsection{Subject-Invariant Decoding}
The primary motivation for the subject-object disentanglement design, introduced in Section \ref{subsec:method_so_dis}, is to decompose the entangled neural feature $\mathbf{F}$ into subject-specific and object-oriented components $\mathbf{Z}_\text{subj}$ and $\mathbf{Z}_\text{obj}$ for a better biometric and semantic decoding. This approach expects object-wise tokens' contribution in $\mathbf{Z}_\text{obj}$ to remain consistent across subjects. Using the object \textit{chair} as an example, we visualize 280 tokens' activations averaged across all correct predictions by subject in Figure \ref{fig:tokens}. It turns out that at the feature level, our method successfully achieves subject-invariant decoding, as token activations display high similarity with only negligible subject-level variations. This outcome demonstrates our model's effectiveness in extracting object-specific information from complex fMRI data, offering a robust framework for multi-subject fMRI decoding.

\begin{figure}[t]
  \centering
  \includegraphics[width=1\textwidth]{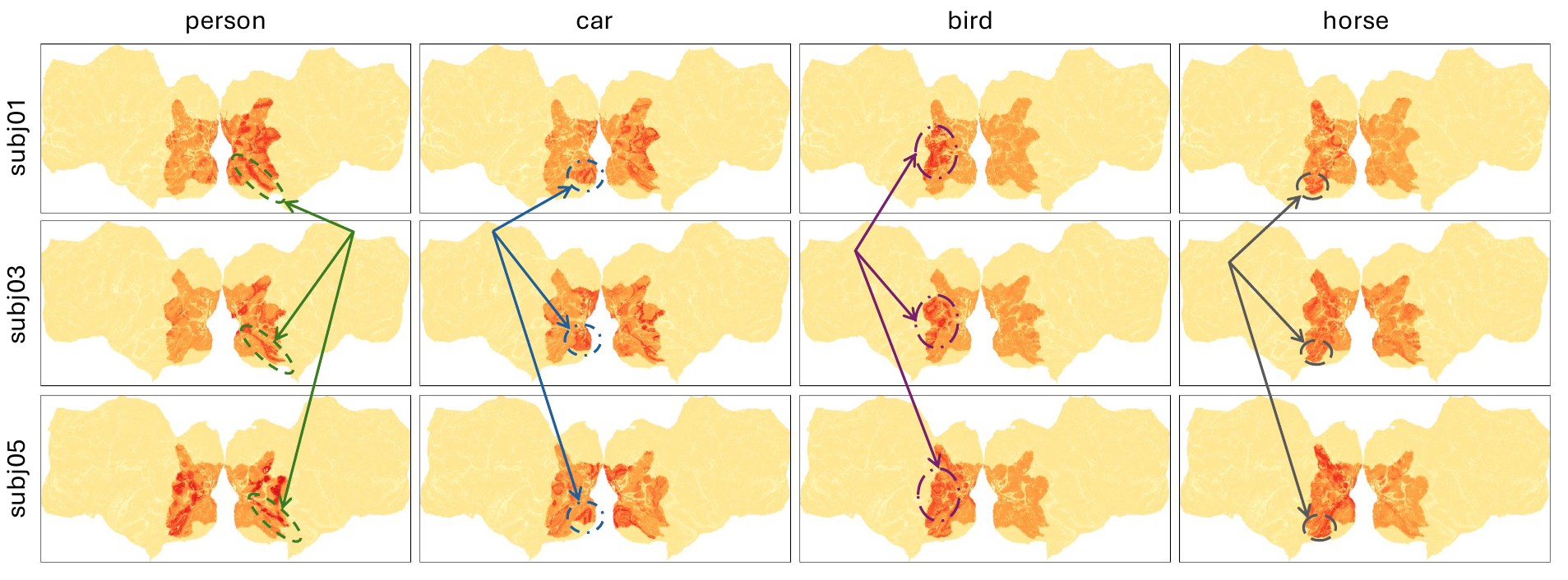}
  \vspace{-5mm}\caption{3D Object-Voxel activations of \textit{person}, \textit{car}, \textit{bird}, and \textit{horse} for subj01, subj03, and subj05.}\vspace{-5mm}\label{fig:vo_3d_mean}
\end{figure}

\subsection{Visual-Neural Relationship}

We empirically investigate the relationship between brain activities and semantic objects in visual stimuli, leveraging both GradCAM \citep{gradcam} and Attention Roll-out \citep{attn_rollout}. 

\noindent\textbf{Subject-wise 1D Activation Pattern}. Taking \textbf{subj01} as an example, we calculate voxel-wise activations within low-level visual regions of interest (ROIs) (V1-V4) and a wider high-level visual ROI in response to three objects: \textit{person}, \textit{horse}, and \textit{chair}. This calculation takes the median activation across all true positive samples predicted by our model on the testset. As shown in Figure \ref{fig:vo_median}, the y-axis represents median activation, while the x-axis represents voxel IDs, providing a clear overview of the active voxels across both low-level and high-level visual ROIs when \textbf{subj01} recognizes each object. The unique activation patterns, with specific voxels responding to each object, indicate that brain voxels function differently from image pixels. In images, object locations are believed spatially random, so evaluating pixel-wise activation does not yield consistent spatial activation patterns. In contrast, fMRI voxels exhibit specialized roles in processing visual information, suggesting that brain voxels are organized by functional responsibility with a degree of spatial invariance—especially in high-level visual ROIs. This conclusion aligns with the existing studies from neuroscience \cite{brain_semantic_selectivity1,brain_semantic_selectivity2}.

\begin{figure}[t]
  \centering
  \includegraphics[width=1\textwidth]{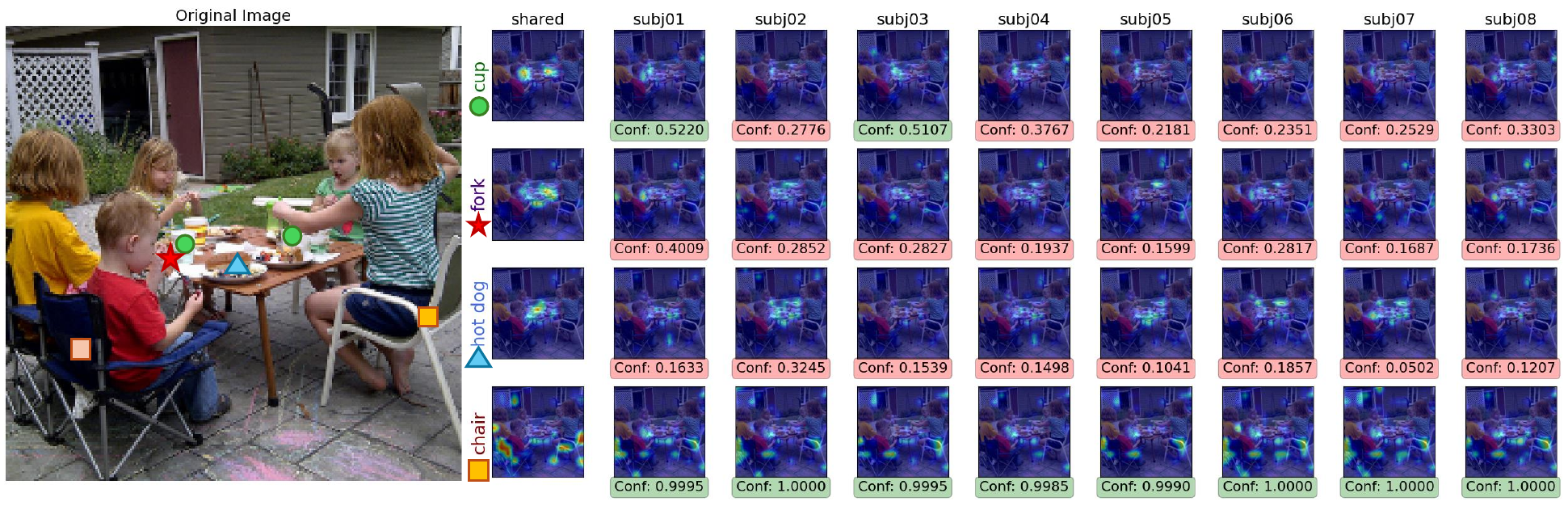}
  \vspace{-5mm}\caption{Variations in subjects' attention to different objects. The leftmost image shows the visual stimulus, labeled with six objects: \textit{person}, \textit{dining table}, \textit{cup}, \textit{fork}, \textit{hot dog}, and \textit{chair}. Four of them are selected for visualization. Plots in the second column represent the shared attention across all subjects, and the remaining eight columns show the residual, subject-specific attention alongside predicted probabilities to compare recognition confidence and priority. 
 }\vspace{-5mm}\label{fig:subj_attn}
\end{figure}

\noindent\textbf{Cross-subject 3D Activation Pattern}. We present 3D brain activation patterns in Figure \ref{fig:vo_3d_mean} for four objects—\textit{person}, \textit{car}, \textit{bird}, \textit{horse}—visualized across three subjects: subj01, subj03, and subj05. Light yellow regions denote non-visual areas that were excluded from the dataset, resulting in uniformly absent signals in these regions. Based on the visualization, the following observations are made:
\begin{itemize}[leftmargin=*]
    \item \textbf{Consistency across subjects}: the objects \textit{bird} and \textit{horse} show a broad similarity across subjects, particularly in the region of the higher visual cortex and predominantly in the left hemisphere. This consistency suggests that certain high-level features associated with animals may be processed in similar ways across individuals, reflecting stable visual processing pathways in the brain.
    \item \textbf{Object sensitivity}: the activation intensity for object \textit{person} appears stronger and more concentrated, indicating that the brain may allocate increased neural resources or ``attention'' to socially relevant stimuli (\textit{people}), compared to less socially significant objects like \textit{bird}. This result is supported by neuroscience research \citep{brain_people1, brain_people2}.
    \item \textbf{Representational flexibility}: while general patterns are shared across subjects, the intensity and spatial distribution of activation vary slightly for certain objects, such as $\textit{car}$. These variations may reflect individual differences in brain anatomy or prior experiences that influence object representation and visual information processing. This flexibility of the brain's adaptability to personal needs and experiences is known as neural plasticity \citep{brain_region1, brain_region2, brain_plasticity}.
\end{itemize}

\subsection{Variations in Subject Attention} 

\begin{wrapfigure}[17]{r}{0.63\textwidth}
\centering\vspace{-4mm}
  \includegraphics[width=.63\textwidth]{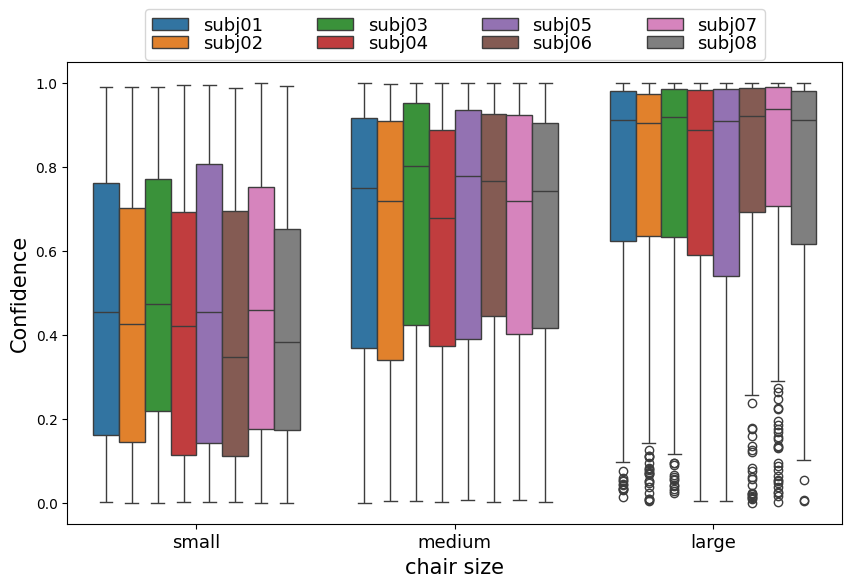}
  \vspace{-6mm}\caption{Recognition of \textit{chair} by object size.}\label{fig:chair_cond}
\end{wrapfigure}

A key contribution of $i$MIND is its use of subject-invariant CLIP visual features to explore how different subjects focus on distinct objects when receiving the same stimulus. Figure \ref{fig:subj_attn} illustrates this attention variation: the first column displays the original visual stimulus with six ground-truth annotations, the second column represents the shared attention map common across all subjects, and the remaining eight columns show the residual, subject-specific attention patterns. Four object-specific attention maps are visualized on rows with predicted logits to compare recognition confidence. 

\begin{wrapfigure}[12]{r}{0.5\textwidth}
  \centering\vspace{-5mm}
  \includegraphics[width=.5\textwidth]{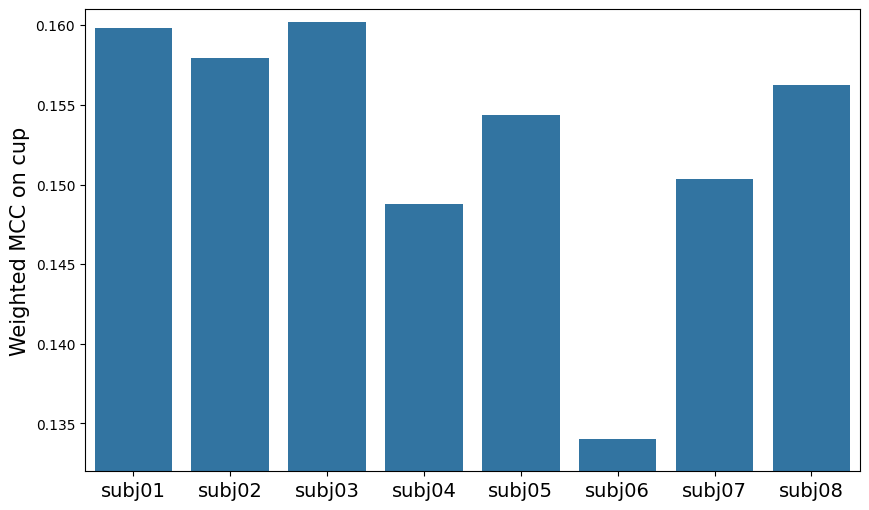}
  \vspace{-6mm}\caption{Sensitivity to \textit{cup} accross subjects.}\vspace{-3mm}\label{fig:cup_mcc}
\end{wrapfigure}

Considering images are shown for only 3s \citep{nsd}, patterns of attention and object recognition offer even more intriguing insights into rapid, automatic processes of visual information and neural encoding:
\begin{itemize}[leftmargin=*]
    \item \textbf{Temporal Constraints on Attention Allocation}: Within a brief 3-second viewing window, the brain must rapidly parse and prioritize elements of a complex scene. Notably, despite occupying only a modest portion of the image, the object \textit{chair} consistently receives high attention across all subjects. This suggests that chairs are processed in an early, feed-forward manner—likely due to their high salience and distinctive visual features that enable rapid recognition under time constraints. To validate this, we analyzed prediction confidence for \textit{chair} in the training set across subjects, grouped by object size (Figure~\ref{fig:chair_cond}). The results confirm that chairs of sufficient size are reliably identified by all participants. This immediate and confident response highlights the efficiency of the visual system in detecting familiar, contextually relevant objects with minimal cognitive effort. Details on Figure~\ref{fig:chair_cond} can be found in \ref{subsec:chair_cond_details}.
    \item \textbf{Subject-specific Focus Under Time Constraints}: Under these brief viewing conditions, subject-specific differences in attention to objects like \textit{cup}, \textit{fork}, and \textit{hot dog} become especially revealing. Variation in attention to \textit{cup}, particularly with subj01 and subj03 achieving recognition confidence (predicted probabilities > 0.5) by focusing more precisely on its location, suggests that these individuals may possess faster or more selective attentional strategies. Such patterns point to individual differences in visual processing speed, attentional control, or perceptual expertise that influence object prioritization under time constraints. These findings are consistent with training results (Figure~\ref{fig:cup_mcc}), where subj01 and subj03 show the highest sensitivity to cups, as measured by weighted MCC. Interestingly, although none of the subjects successfully recognize \textit{fork} or \textit{hot dog} in Figure~\ref{fig:subj_attn}, all but subj02 allocate more attention to the \textit{fork} than the \textit{hot dog}, suggesting a subtle yet consistent attentional bias that may reflect object familiarity or contextual relevance.
\end{itemize}

To our best knowledge, the proposed $i$MIND is the first model to capture subtle variations in how quickly and differently individuals allocate attention within a constrained time frame, demonstrating the model's robustness in simulating real-world neural processes. The model's ability to account for both shared and individual-specific attention patterns in response to brief stimulus exposure can inform the development of neural decoding approaches that better reflect human variability, especially in time-sensitive applications like real-time scene analysis or autonomous driving. Complete details for all visualized figures are provided in the Appendix.

In sum, the fact that subjects allocate attention differently within just a few seconds underscores the efficiency of neural mechanisms in prioritizing objects and the role of individual cognitive differences. This rapid, nuanced attention mapping highlights how our $i$MIND framework captures the interplay of shared and individual neural patterns, bridging cognitive neuroscience with computational modeling to decode visual attention in real-world scenarios.

\begin{wraptable}[12]{r}{0.43\textwidth}  
\centering\vspace{-4mm}
    \caption{Ablation on loss functions.}\vspace{-3mm}\label{tb:loss}
    \rowcolors{2}{rowgray}{white}
    \begin{tabular}{c|cccc}
    \toprule
     ID            & $\mathcal{L}_\text{subj}$ & $\mathcal{L}_\text{orth}$ & $\lambda$    & mAP (\%)\\ \midrule
    \textbf{Full} & \checkmark                & \checkmark                & \textbf{.1}  & \textbf{78.36}  \\\midrule
    1             & \checkmark                &                           &              & -7.15 ($\downarrow$)           \\
    2             &                           & \checkmark                & \textbf{.1}           & -11.17 ($\downarrow$)          \\
    3             &                           &                           &              & -11.07 ($\downarrow$)          \\\midrule
    4             & \checkmark                & \checkmark                & \textbf{.01} & -0.94 ($\downarrow$)           \\
    5             & \checkmark                & \checkmark                & \textbf{1}   & -2.95  ($\downarrow$)          \\ 
    6             & \checkmark                & \checkmark                & \textbf{10}  & -0.84  ($\downarrow$)          \\
    \bottomrule
    \end{tabular}
\end{wraptable}

\subsection{Ablation Study}\label{subsec:ablation}
Our novel designs—subject-object disentanglement and the dual-decoding framework—are considered two key factors in achieving SOTA semantic decoding performance. To evaluate their effectiveness and necessity, we test combinations of the two loss functions, $\mathcal{L}_\text{subj}$ and $\mathcal{L}_\text{orth}$, along with a trade-off hyperparameter $\lambda$. Table \ref{tb:loss} confirms that both $\mathcal{L}_\text{orth}$ for subject-object disentanglement and the dual-decoding design are crucial for achieving high semantic performance. Moreover, the trade-off parameter seems to have a minimal effect on the overall results. The optimal model utilizes all three loss functions with a trade-off parameter of $\lambda=0.1$. More studies on hyperparameters' impacts on semantic decoding are provided in Appendix \ref{subsec:ablation_hd}.

%% file: sec/M5_discussion.tex
\section{Conclusion}
\label{sec:disscussion}

In this paper, we introduce an innovative multi-subject dual-decoding framework that decomposes latent fMRI representations into distinct subject-specific and object-specific components using a robust basis transformation. This approach enables precise biometric decoding through individualized neural features, while shared object-oriented features facilitate subject-invariant semantic decoding by querying with CLIP-derived visual representations. Our framework not only establishes a new benchmark for semantic decoding accuracy but also reveals variations in attentional focus across subjects when viewing identical visual stimuli. Additionally, we construct object-specific activation patterns at the voxel level, offering data-driven insights into the brain's visual processing mechanisms. 

In future work, we aim to leverage large-scale fMRI datasets to develop more robust and informative pretrained models for extracting latent neural features. Additionally, we plan to collaborate with brain scientists to deepen our understanding of how specific voxel patterns in fMRI data relate to semantic object representations. This domain knowledge will help bridge the gap between visual features and neural signals, further enhancing the interpretability and accuracy of brain-based decoding models.



\clearpage

%% file: sec/A1_appendix.tex
\section{Technical Appendices and Supplementary Material}

\subsection{Related Work}
\label{subsec:realte_works}

\subsubsection{Single-Subject v.s. Multi-Subject Models}
The application of deep learning to neural data has initially centered on single-subject models tailored to individual participants. BrainDiVE \citep{brain-dive} adopts a generative approach, synthesizing images predicted to activate specific regions of the human visual cortex. Moving beyond visual modalities, Mind Reader \citep{mindreader} incorporates textual information to reconstruct complex images containing multiple objects from brain activities. Extending this multimodal approach, BrainSCUBA \citep{brain-scuba} takes advantage of contrastive vision-language models and large-language models to generate voxel-wise captions, eliminating the need for human-annotated voxel-caption data. While single-subject models have achieved notable success, they face inherent limitations. These models require large amounts of subject-specific data to train robust models, which is challenging given the high costs and effort involved in collecting fMRI data. Furthermore, they are prone to overfitting, exhibit poor generalizability across individuals, and struggle with scalability when applied to larger datasets or diverse populations.

To overcome these challenges, multi-subject models aim to unify data across participants, enabling shared representation learning. However, this approach introduces significant complexity due to inter-subject variability, which arises from static anatomical differences and dynamic functional responses. Various methods have been proposed to overcome these obstacles. \cite{emb} employs subject embeddings and recurrent architectures to account for inter-trial and inter-subject variability, outperforming many single-subject models in predicting MEG time series. MindBridge \citep{mind-bridge}  introduces a biologically inspired aggregation function and a cyclic fMRI reconstruction mechanism to achieve subject-invariant representation learning. MindEye2 \citep{mindeye2} aligns spatial patterns of fMRI activity to a shared latent space using subject-specific ridge regression, improving out-of-subject generalization with limited training data and achieving state-of-the-art results in image retrieval and reconstruction. More recently, CLIP-MUSED \citep{clip_mused} introduced learnable subject-specific tokens to facilitate the aggregation of multi-subject data without a linear increase in model parameters. This approach integrates representational similarity analysis (RSA) to guide token representation learning based on the topological relationships of visual stimuli in the latent visual space. 

These advancements demonstrate the potential of multi-subject models to surpass the limitations of single-subject approaches, providing more generalizable and scalable solutions for neural decoding tasks. However, to the best of our knowledge, existing multi-subject neural decoding models predominantly adopt what we term a \textbf{suppressive} strategy for handling inter-subject variability. This approach aims to minimize subject-specific differences during learning, progressively refining features to become more task-relevant as the model deepens. In these frameworks, subject-specific information is often treated as noise or an obstacle to effective decoding. In contrast, our $i$MIND framework proposes an \textbf{instructive} strategy. Rather than suppressing subject-specific differences, our model embraces this variability by explicitly disentangling subject-specific features from task-relevant ones. By doing so, $i$MIND not only preserves individual-specific neural representations but also leverages them positively to enhance both subject-specific and shared task-related decoding. This dual-decoding approach enables $i$MIND to achieve superior performance while offering insights into both individual neural patterns and shared semantic representations.

\subsubsection{Vision-Neural Interactions}
Decoding visual information from neural signals is an inherently multi-modal task, involving the alignment and interaction of at least two modalities: images and neural signals (eg., fMRI). Broadly speaking, approaches for vision-neural modality alignment can be categorized into two branches based on the direction of projection between visual and neural spaces.

The first branch projects neural signals into a pre-trained latent visual space. This approach is exemplified by works such as \cite{mindeye1}, which maps flattened spatial patterns of fMRI activity across 3D cortical tissue cubes into the image embedding space of a pre-trained CLIP model. Similarly, \cite{data_prep2} predicts latent representations of presented images from fMRI signals within the early visual cortex. Other notable works in this branch include \cite{mind-bridge, mind-vis, mindreader, mindeye2}, which leverage pre-trained visual generative models, such as GANs \citep{rec_by_gan2} and diffusion models, for reconstruction tasks. These models capitalize on large-scale visual datasets and avoid re-training resource-intensive generative architectures, which would otherwise be infeasible given the scarcity of paired neural-visual data. The second branch adopts the opposite approach by projecting latent visual image features into the neural space. This method is particularly useful for generating synthetic stimuli that activate specific brain regions, enabling the study of feature preferences in different areas of the brain. Classic examples include \cite{brain-dive, brain-scuba, i2fmri3}, which investigate neural activation patterns in response to synthetic stimuli derived from visual features.

Our $i$MIND model takes a fundamentally different approach to vision-neural modality interaction. Rather than projecting between modalities, we use CLIP-derived vision features as queries to extract corresponding neural object features directly from neural representations. This design choice is motivated by several factors. First, as a semantic neural decoding framework, $i$MIND does not rely on resource-intensive generative models. Second, direct projections between modalities often result in significant information loss and modality gaps that require careful handling. Most importantly, our approach is expected to enable the investigation of subject-specific attention variations when viewing the same visual stimuli. By treating the CLIP vision features as pseudo-ground-truths for object presence, we leverage their subject-invariant properties as an anchor to explore how neural responses to specific objects differ across subjects. This design uniquely aligns with the goals of understanding inter-subject variability in neural decoding.

\subsection{Theoretic Validation for Basis Transformation}
\label{subsec:bases_proof}
In this section, we present the basis transformation in linear algebra and establish the relationship between coordinates in different bases. The derivation ensures clarity in transitioning from one basis to another, essential for interpreting subject-specific neural space $\mathcal{F}_\text{subj}$ and object-specific neural space $\mathcal{F}_\text{obj}$ mentioned at the end of Section \ref{subsec:method_so_dis} in the main paper. We begin with the following formal claim:

\paragraph{Claim}
Let $\mathcal{V}$ be a vector space of cardinality $d$ over $\mathbb{R}$, with a standard basis $\mathbf{E}=\{\mathbf{e}_1,\mathbf{e}_2,\dots,\mathbf{e}_d\}$ of $\mathbb{R}^d$ and an arbitrary basis $\mathbf{B}=\{\mathbf{b}_1,\mathbf{b}_2,\dots,\mathbf{b}_d\}$ of $\mathbb{R}^d$. For any vector $\mathbf{v}\in\mathcal{V}$, if its coordinate with respect to the standard basis $\mathbf{E}$ is given by $[\mathbf{v}]_\mathbf{E}\in\mathbb{R}^d$, then its coordinate with respect to the basis $\mathbf{B}$ can be derived as:
\begin{align}
    \mathbf{[v]_B=P_{E\rightarrow B}\cdot[v]_E},
\end{align}
where $\mathbf{P_{E\rightarrow B}}\in\mathbb{R}^{d\times d}$ is the \textit{change-of-basis} matrix from $\mathbf{E}$ to $\mathbf{B}$, defined as:
\begin{align}
    \mathbf{P_{E\rightarrow B}}:=\begin{bmatrix} 
| & | & \cdots & | \\ 
\mathbf{b}_1 & \mathbf{b}_2 & \cdots & \mathbf{b}_d \\ 
| & | & \cdots & | 
\end{bmatrix}^{-1}.
\end{align}

\paragraph{Proof}
Since $\mathbf{v}\in\mathcal{V}$ and $\mathbf{E}$ forms a basis for the vector space $\mathcal{V}$, $\mathbf{v}$ can be written as a linear combination of all basis vectors from $\mathbf{E}$:
\begin{align}\label{eq:linear_e}
    \mathbf{v}=\sum_{i=1}^da_i\mathbf{e}_i,
\end{align}
where $a_i\in\mathbb{R}$ are scalars. In this case, the coordinate of $\mathbf{v}$ with respect to $\mathbf{E}$ is:
\begin{align}\label{eq:cor_e}
    \mathbf{[v]_E}=(a_1,a_2,\dots,a_d)^\top\in\mathbb{R}^d.
\end{align}
Similarly, because $\mathbf{B}$ also forms a basis for $\mathcal{V}$, we can express $\mathbf{v}$ as:
\begin{align}\label{eq:linear_b}
    \mathbf{v}=\sum_{i=1}^dw_i\mathbf{b}_i,
\end{align}
where $w_i\in\mathbb{R}$ are scalars. The coordinate of $\mathbf{v}$ with respect to $\mathbf{B}$ is:
\begin{align}\label{eq:cor_b}
    \mathbf{[v]_B}=(w_1,w_2,\dots,w_d)^\top\in\mathbb{R}^d.
\end{align}
Since each basis vector $\mathbf{e}_i$ within $\mathbf{E}$ is also an element of the vector space $\mathcal{V}$, it can also be written as a linear combination of all basis vectors from $\mathbf{B}$:
\begin{align}\label{eq:linear_eb}
    \mathbf{e}_i=\sum_{j=1}^dp_{ji}\mathbf{b}_j.
\end{align}
Writing the equation above in matrix form, we obtain:
\begin{align}\label{eq:relation_eb}
    \begin{bmatrix} 
    | & | & \cdots & | \\ 
    \mathbf{e}_1 & \mathbf{e}_2 & \cdots & \mathbf{e}_d \\ 
    | & | & \cdots & | 
    \end{bmatrix}&=\begin{bmatrix} 
    p_{11} & p_{12} & \cdots & p_{1d} \\ 
    p_{21} & \ddots &        & p_{2d} \\ 
    \vdots &        & \ddots & \vdots \\ 
    p_{d1} & p_{d2} & \cdots & p_{dd}
    \end{bmatrix}\begin{bmatrix} 
    | & | & \cdots & | \\ 
    \mathbf{b}_1 & \mathbf{b}_2 & \cdots & \mathbf{b}_d \\ 
    | & | & \cdots & |
    \end{bmatrix}.
    \end{align}
Denoting the middle matrix as $\mathbf{P}$ and solving it, we get:
    \begin{align}\label{eq:matrix_p}    
    \mathbf{P}:=
    \begin{bmatrix} 
    p_{11} & p_{12} & \cdots & p_{1d} \\ 
    p_{21} & \ddots &        & p_{2d} \\ 
    \vdots &        & \ddots & \vdots \\ 
    p_{d1} & p_{d2} & \cdots & p_{dd}
    \end{bmatrix}&=\begin{bmatrix} 
    | & | & \cdots & | \\ 
    \mathbf{b}_1 & \mathbf{b}_2 & \cdots & \mathbf{b}_d \\ 
    | & | & \cdots & | 
    \end{bmatrix}^{-1}.
\end{align}
Next, let's plug each $\mathbf{e}_i$ into Eq. \eqref{eq:linear_e} using the formulation of Eq. \eqref{eq:linear_eb}:
\begin{align}
    \mathbf{v} &= \sum_{i=1}^da_i\mathbf{e}_i 
    =\sum_{i=1}^d(a_i\sum_{j=1}^dp_{ji}\mathbf{b}_j) 
    =\sum_{j=1}^d(\sum_{i=1}^da_ip_{ji})\mathbf{b}_j.
\end{align}
Combining with Eq. \eqref{eq:linear_b}, we have:
\begin{align}
    \sum_{j=1}^dw_j\mathbf{b}_j=\sum_{j=1}^d(\sum_{i=1}^da_ip_{ji})\mathbf{b}_j.
\end{align}
Move everything to the left hand:
\begin{align}\label{eq:l_ind}
    \sum_{j=1}^d[w_j-(\sum_{i=1}^da_ip_{ji})]\mathbf{b}_j=0.
\end{align}
According to the claim, $\mathbf{B}=\{\mathbf{b}_1,\mathbf{b}_2,\dots,\mathbf{b}_d\}$ is a basis of $\mathbf{R}^d$. Therefore, Eq. \eqref{eq:l_ind} holds if and only if:
\begin{align}
    w_j-\sum_{i=1}^da_ip_{ji}=0\quad\text{for } j=1,2\dots,d.
\end{align}
Equivalently, we have:
\begin{align}
    \begin{bmatrix} 
    w_1   \\ 
    w_2   \\ 
    \vdots\\ 
    w_d 
    \end{bmatrix}=\sum_{i=1}^da_i\begin{bmatrix} 
    p_{1i} \\ 
    p_{2i} \\ 
    \vdots \\ 
    p_{di} 
    \end{bmatrix}=\begin{bmatrix} 
    p_{11} & p_{12} & \cdots & p_{1d} \\ 
    p_{21} & \ddots &        & p_{2d} \\ 
    \vdots &        & \ddots & \vdots \\ 
    p_{d1} & p_{d2} & \cdots & p_{dd}
    \end{bmatrix}\begin{bmatrix} 
    a_1 \\ 
    a_2 \\ 
    \vdots \\ 
    a_d
    \end{bmatrix}.
\end{align}
Using the coordinates expression defined in Eq. \eqref{eq:cor_e} and Eq. \eqref{eq:cor_b} along with Eq. \eqref{eq:matrix_p}, we obtain the following equation:
\begin{align}
    \mathbf{[v]_B=P\cdot [v]_E}\quad\text{where  }\mathbf{P}=\begin{bmatrix} 
    | & | & \cdots & | \\ 
    \mathbf{b}_1 & \mathbf{b}_2 & \cdots & \mathbf{b}_d \\ 
    | & | & \cdots & | 
    \end{bmatrix}^{-1}.
\end{align}
Finally, we complete the proof of the claim.

In our $i$MIND model, subject-object disentanglement is achieved through a basis transformation described above, where the new basis $\mathbf{B}$ is treated as learnable parameters optimized by loss propagation. We further enforce the orthonormality constraint on $\mathbf{B}$, as this ensures that any subspace spanned by a subset of $\mathbf{B}$ is orthogonal (complementary) to its counterparts in the original feature space. Specifically, in our model, this constraint guarantees that the subject-specific neural space $\mathcal{F}_\text{subj}$, spanned by $\mathbf{B}_\text{subj}$, and the object-specific neural space $\mathcal{F}_\text{obj}$, spanned by $\mathbf{B}_\text{obj}$, are complementary and non-overlapping. Consequently, this orthogonal decomposition yields a perfect separation of subject-specific and object-specific neural features within the latent neural representations.

\subsection{NSD dataset, Pre-processing, and Implementation }
\subsubsection{NSD Dataset} 
The Natural Scenes Dataset (NSD) \citep{nsd} is a groundbreaking resource in cognitive neuroscience and artificial intelligence, designed to capture extensive, high-resolution fMRI data during natural scene perception. It includes whole-brain fMRI measurements of eight human participants at 7T field strength, with a spatial resolution of 1.8 mm. Participants viewed a total of $70,566$ natural scene images, with $10,000$ unique images per subject (9,000 unique to each participant and $1,000$ shared across all participants). Images were sourced from the richly annotated Microsoft COCO dataset \citep{mscoco}, ensuring ecological relevance and diversity. The experiment was conducted over 30–40 sessions per participant, taking a rapid event-related design with continuous recognition tasks to guarantee engagement and probe both short- and long-term memory processes. During neural recording, participants were tasked with identifying objects in the image, with each visual stimulus presented for only three seconds per trial. This design makes the NSD particularly well-suited for investigating the mechanisms of rapid attention and visual recognition in human vision. Advanced pre-processing completed by the authors, including denoising and voxel-specific hemodynamic response modeling, yielded high-quality single-trial beta estimates with exceptional signal-to-noise ratios. Complementing the functional data, NSD includes extensive anatomical scans, resting-state data, and behavioral performance measures, enabling multi-faceted investigations of vision and memory. This dataset, with its unparalleled scale and quality, serves as a valuable benchmark for developing and testing machine learning models that aim to decode brain activity and simulate neural representations of natural scenes. In addition, the NSD dataset supports multiple widely used neuroimaging atlases to facilitate data analysis and integration with existing frameworks. Functional data are provided in both native cortical surface space and standard volumetric spaces, including fsaverage \citep{fsaverage} and MNI152 \citep{mni152}, enabling compatibility with tools like FreeSurfer \citep{freesurfer} and FMRIB Software Library. Additionally, the dataset includes manually defined regions of interest (ROIs) for retinotopic mapping and category-selective areas, such as the early visual cortex and higher-order regions in the ventral visual stream, which is the atlas that we used in our $i$MIND model. These comprehensive atlases allow researchers to seamlessly apply NSD data to diverse analytic pipelines and cross-study comparisons.

\subsubsection{Preprocessing} 
\label{subsec:preprocess}
Our preprocessing pipeline begins with splitting the dataset into training and testing sets. Due to incomplete sessions and data availability constraints, not all trials are accessible for every subject, resulting in a total of 213,000 trials across all participants. Among these, neural recordings corresponding to 1,000 images viewed by all subjects are allocated to the testing set, comprising a total of 21,118 test trials. The remaining neural recordings, corresponding to images viewed exclusively by individual subjects, are included in the training set, resulting in 191,882 training trials. Both training and testing trials are standardized voxel-wise using the mean and standard deviation calculated from the training set. Since each image is presented to a subject three times, we average the fMRI responses across repetitions for each image within each subject. This results in 69,566 training samples and 7,674 testing samples, allowing us to train and evaluate a single multi-subject model across all eight subjects. For each sample, we use the \textit{nsdgeneral} atlas provided by the NSD dataset to extract visual voxel signals as a 1D vector. However, the number of visual voxels varies between subjects due to anatomical differences, with the voxel length $L_s$ ranging from $12,682$ to $17,907$ across subjects.  To unify the input length, we apply a padding strategy inspired by Mind-Vis \citep{mind-vis}, which conducts wrap-around padding. This approach avoids issues arising from truncation or constant padding to the maximum voxel length. Additionally, since our fMRI encoder is based on a Vision Transformer (ViT), which requires input lengths divisible by the user-defined patch size (64 in our model), we adjust the uniform voxel length $L$ accordingly. The final voxel length across subjects is set to $L=17,920$, ensuring compatibility with the model while maintaining consistency across participants.

\subsubsection{Implementation Details}
\label{subsec:implementation}
As described in Section \ref{sec:method} of the main paper, our proposed architecture is trained in two stages. The first stage involves pre-training a ViT-based masked autoencoder, similar to SC-MBM \citep{mind-vis}, using a self-supervised fMRI reconstruction task. In this stage, we choose a patch size of 64 voxels with a masking ratio of 0.75. The encoder has a hidden dimension of 768 and consists of 12 layers of 6-head self-attention, while the decoder has a hidden dimension of 512 and 8 layers of 8-head self-attention. In the second stage, we discard the decoder and inherit only the encoder from the first stage, which outputs pre-trained feature $\mathbf{F}\in\mathbb{R}^{N\times d}$ with $N=280$ and $d=768$. We set the object neural space dimension $d_\text{obj} = 700$ and choose a 4-head cross-attention module for fMRI-vision feature interactions. The CLIP visual encoder we used is clip-vit-base-patch16 released by OpenAI, which remains frozen at all stages of the proposed framework. A trade-off parameter $\lambda$ of 0.1 is set by default to enforce the orthonormal constraint of the learnable basis $\mathbf{B}$. A detailed investigation is provided in Section \ref{subsec:ablation} and Appendix \ref{subsec:ablation_hd}. During this stage, all parameters are optimized end-to-end for subject and object classification tasks. For either stage, we train the model for 100 epochs, including 10 warm-up epochs. The learning rate is initialized at $7.5 \times 10^{-4}$ and terminated at zero adjusted dynamically by a cosine scheduler. The batch size is set to 200. Optimization is performed via the AdamW optimizer with a weight decay of 0.05. All experiments are conducted on two Nvidia RTX 6000 Ada GPUs, with the first stage taking approximately 1.5 hours and the second stage around 2 hours to complete.

\subsection{Subject Classification Baselines}
\label{subsec:subj_baseline}
To the best of our knowledge, no existing models support subject classification. To provide a comprehensive evaluation, we established baseline models using the exact fMRI preprocessing steps in Appendix \ref{subsec:preprocess} and conducted both supervised and unsupervised biometric decoding.All methods are trained and tested on identical data splits and fMRI voxel sets as iMIND, ensuring a fair comparison on the same held-out unseen test set.

For supervised learning, we employ a single linear layer trained in two ways:
\begin{itemize}[leftmargin=*]
    \item Linear Regression: We minimize the mean squared error (MSE) between the input (padded fMRI voxel signals) and the target (one-hot subject IDs), using the ordinary least squares closed-form solution;
    \item Classification: We train an identical architecture with cross-entropy loss, treating subject identification as a standard classification task.    
\end{itemize}

For unsupervised learning, we evaluate K-Means clustering with two distance metrics:
\begin{itemize}[leftmargin=*]
    \item Euclidean (L2) distance;
    \item Cosine similarity.
\end{itemize}
Since the number of subjects is known (8), we set the number of clusters to 8 as well. To measure performance, we compute accuracy and MCC by optimally aligning the learned clusters with ground-truth subject IDs.

\subsection{Subject Generalizability}
For completeness, we conducted an additional experiment to assess subject generalizability within NSD as shown in Table \ref{tb:gen}. In our original setup (denoted as M8), iMIND was trained on data from all 8 subjects. For this experiment, we introduced a variant (M7), where iMIND was trained using only the first 7 subjects and tested on the held-out data of subj08. \textbf{M7/M8} achieved an overall mAP of \textbf{.7904/.7842} on the first 7 subjects and \textbf{.7842/.7909} on subj08. These results demonstrate that our proposed method exhibits a reasonable and strong generalizability in neural signal semantic decoding, particularly for unseen subjects. 
\begin{table}[t]
\centering
\caption{Generalizability to unseen subjects}\label{tb:gen}
\begin{tabular}{c|cccc}
\toprule
ID                                  & Trained on & Tested on   & mAP (\%)  \\ \midrule
\multirow{2}{*}{M7}                 & subj01--07 & subj01--07  & .7904 \\ 
                                    & subj01--07 & subj08      & .7842 \\ \midrule
\multirow{2}{*}{M8 (\textbf{Full})} & subj01--08 & subj01--07  & .7842 \\
                                    & subj01--08 & subj08      & .7909 \\
\bottomrule
\end{tabular}
\end{table}

\subsection{Ablations on Hyper-parameters Affecting Semantic Decoding}
\label{subsec:ablation_hd}
We investigate the impact of two key hyperparameters on the performance of semantic decoding: $d_\text{obj}$, the dimension of the neural object space for subject-object disentanglement in Section \ref{subsec:method_so_dis}, and $h$, the number of heads in multi-head cross attention module in Eq. \eqref{eq:cross_attn}. According to Table \ref{tb:dobj}, increasing the number of heads does not necessarily lead to improved performance, as it may result in potential overfitting. In addition, we found that performance degradation remains minimal as long as there is sufficient feature space allocated for the neural object information. Ultimately, the optimal object classification performance, in terms of mAP, is achieved with $h=4$ and $d_\text{obj}=700$.
\begin{table}[t]
\centering
\caption{Ablation on heads and $d_\text{obj}$.}\label{tb:dobj}
\begin{tabular}{c|cccc}
\toprule
ID            & Head(s)    & $d_\text{obj}$ & mAP (\%) / $+-$ \\ \midrule
\textbf{Full} & \textbf{4} & \textbf{700}   & \textbf{78.36} \\
\midrule
1             & \textbf{1} & 700            & -1.95 \\
2             & \textbf{2} & 700            & -1.12 \\
3             & \textbf{6} & 700            & -1.35 \\
4             & \textbf{8} & 700            & -9.18 \\\midrule
5             & 4          & \textbf{100}   & -4.01 \\
6             & 4          & \textbf{200}   & -2.11 \\ 
7             & 4          & \textbf{300}   & -1.04 \\
8             & 4          & \textbf{400}   & -1.38 \\ 
9             & 4          & \textbf{500}   & -0.89 \\
10            & 4          & \textbf{600}   & -1.14 \\ 
\bottomrule
\end{tabular}
\end{table}

\subsection{Limitations}
\label{subsec:limit}

While our method achieves state-of-the-art performance in both semantic and biometric decoding tasks, several limitations remain unresolved.

First, the current approach to neural feature extraction may not be optimal. Although functional, the pretrained neural reconstruction stage produces fMRI reconstructions—both numerically and visually—that underperform compared to the ground-truth voxel signals. This suggests that the masked autoencoder (MAE) backbone may not be the most effective architecture for this task, warranting further exploration. 

Second, flattening voxel inputs discards crucial spatial relationships among neighboring voxels, despite neuroscientific evidence that proximal voxels exhibit functional coupling in visual processing. Future work could explore advanced architectures—such as 3D SwinTransformers, which are explicitly designed for volumetric fMRI data and have demonstrated efficacy in neurological disease diagnosis—to better preserve spatial hierarchies and improve feature learning.

Last, our analysis of visual mechanisms relies on post hoc interpretation methods (Grad-CAM and Rollout), which provide only approximate explanations of model behavior. A more principled approach would involve explainable-by-design architectures for fMRI feature extraction, which we leave for future work.

\subsection{Broader Impacts}
\label{subsec:impacts}

Our work represents a pioneering step toward decoding the brain's visual processing mechanisms, with far-reaching implications for both neuroscience and artificial intelligence. By modeling how the brain transforms visual signals into neural activity and high-level semantics, we aim to uncover the fine-grained functional organization of visual regions—such as those specialized for distinguishing closely related objects (e.g., dogs vs. cats).

This understanding could enable breakthroughs in brain-computer interfaces (BCIs), where precise neural decoding could restore or augment vision for impaired individuals. Conversely, it also raises ethical considerations: the same principles could theoretically be used to manipulate neural signals, artificially inducing semantic perceptions (e.g., generating "fake" visual concepts in the brain). Such capabilities would necessitate rigorous ethical frameworks to prevent misuse while maximizing societal benefit.

Further, our computational approach bridges AI and neuroscience, offering interpretable models that could inspire more biologically plausible machine vision systems. By aligning artificial and biological vision, we may accelerate progress in both fields—from improving AI's robustness to advancing treatments for neurological disorders.

\subsection{Technical Details on Object-Voxel Visualization}
In our experiments, we empirically investigate the relationship between brain activities and semantic objects in visual stimuli. In this section, we detail how voxel contributions to object recognition are visualized in our framework. Given an input fMRI voxel signal $\mathbf{V}\in\mathbb{R}^L$, we obtain its object-specific neural feature map $\mathbf{Z}_\text{obj}\in\mathbb{R}^{N\times d_\text{obj}}$ before cross attention module in our model. Using GradCAM \citep{gradcam}, we are able to build an activation map $\mathbf{t}\in\mathbb{R}^N$, which quantifies the contribution of each neural token $\mathbf{z}_\text{obj}\in\mathbb{R}^{d_\text{obj}}$ in object-specific neural feature map $\mathbf{Z}_\text{obj}\in\mathbb{R}^{N\times d_\text{obj}}$ to the correct recognition of the object of interest. 

Unlike CNN-based models, which can simply resize $t$ to match the size of the input $\mathbf{V}$ because of their spatial invariance nature (a one-to-one correspondence between input patches and latent feature vectors), our ViT-based neural encoder in $i$MIND model lacks such spatial invariance properties by default. For instance, the first neural token in $\mathbf{Z}_\text{obj}$ does not necessarily correspond to the first voxel patch of $\mathbf{V}$, making direct resizing infeasible.

To address this, we leverage Attention Roll-out \citep{attn_rollout} to approximate how information flows from the input voxels $\mathbf{V}$ to the neural tokens in feature map $\mathbf{Z}_\text{obj}$. Specifically, this information flow in our ViT-based encoder can be measured as follows:
\begin{align}\label{eq:attn_rollout}
    \mathbf{A}^{l}=\mathbf{A}^{(l)}\mathbf{A}^{(l-1)}\cdots\mathbf{A}^{(2)}\mathbf{A}^{(1)},
\end{align}
where $\mathbf{A}^{(k)}\in\mathbb{R}^{N\times N}$ represents the attention weights in the $k$-th self attention layer of the ViT encoder. Based on the mathematical property of the self-attention mechanism, the element $a_{ij}$ of the attention weights $\mathbf{A}$ at every transformer block defines how much attention flows from the token $j$ in the previous layer to the token $i$ in the next layer. Therefore, each element $a_{ij}^l$ within $\mathbf{A}^l$ defined in Eq. \eqref{eq:attn_rollout} quantifies the degree of information flow from the $j$-th voxel patch of input fMRI signal $\mathbf{V}\in\mathbb{R}^L$ to the $i-$th token in the feature map $\mathbf{Z}_\text{obj}$. 

Next, we combine the GradCAM-based token contributions $\mathbf{t}\in\mathbb{R}^N$ with the cumulative information flow $\mathbf{A}^l$ to derive the voxel-level activation measurement $\mathbf{T}$:
\begin{align}
    \mathbf{T}=\mathbf{t}\cdot\mathbf{A}^l\in\mathbb{R}^{N}.
\end{align}
Here, $\mathbf{T}$ measures the contribution of each voxel path immediately after embedding the original 1D voxel signal $\mathbf{V}$ of length $L$ into $N$ patches. Since $\mathbf{T}$ is now positionally aligned with the input voxel patches, it can be safely upsampled from size $N$ to $L$ to obtain a voxel-level activation map for each fMRI sample. Unfortunately, this upsampled activation map is partially synthetic because wrap-around padding was applied during preprocessing to achieve a uniform, model-compatible voxel length $L$ across subjects. This padding introduces artificial ``fake'' voxels. The advantage of the wrap-around padding strategy is that it allows us to trace the origins of these fake voxels. To restore the true voxel activation map, we retain the activations of the real voxels, while for the fake voxels, we trace their origins and assign their values as the maximum of the original and artificial activations. This approach ensures that the restored activation map accurately represents the contributions of real voxels while mitigating the impact of synthetic padding.

Finally, this process is repeated across all samples containing the object of interest, enabling us to investigate voxel semantic selectivity, as illustrated in Figure \ref{fig:vo_median} of the main paper. To achieve the 3D activation like Figure \ref{fig:vo_3d_mean} in the main paper, we can just map each flattened voxel back to 3D brain space using the provided \textit{nsdgeneral} atlas.  This methodology allows us to map neural activations back to their voxel-level origins, providing insights into the relationship between neural representations and object recognition.

\subsection{Technical Details on Figure \ref{fig:chair_cond}}\label{subsec:chair_cond_details}
We first computed the pixel occupation ratio for all training images containing the object \textit{chair}. This ratio was derived by dividing the number of chair pixels (using MSCOCO annotations for masking) by the total image resolution. Since the raw pixel ratios exhibited a highly skewed distribution, we applied a log transformation to approximate a normal-like distribution as shown in Figure \ref{fig:chair_dist}.
\begin{figure}[h]
  \centering
  \includegraphics[width=1\textwidth]{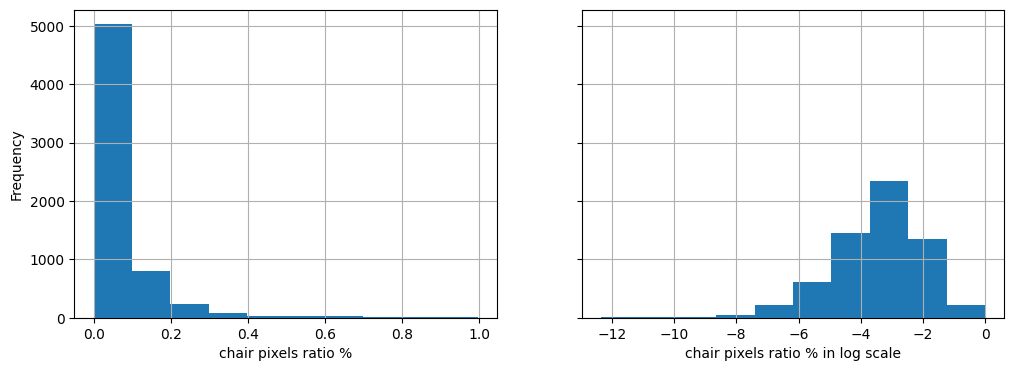}
  \caption{\textit{Chair} pixel ratio distribution in training set}
\end{figure}\label{fig:chair_dist}

Next, we calculated the mean $\mu$ and standard deviation $\sigma$ of the log-transformed ratios. To partition the chairs into size-based categories, we defined three intervals:
\begin{itemize}[leftmargin=*]
    \item Small chairs: $(-\infty,\mu-0.5\sigma)$
    \item Medium chairs: $(\mu-0.5\sigma,\mu+0.5\sigma)$
    \item Large chairs: $(\mu+0.5\sigma, 0)$
\end{itemize}

Finally, for each subject, we generated a boxplot of the predicted probabilities for the \textit{chair} class, stratified by these size groups.

\subsection{Technical Details on Figure \ref{fig:cup_mcc}}\label{subsec:cup_mcc_details}
To analyze differences in subjects' sensitivity to the object \textit{cup}, we first computed the pixel occupation ratio for all training images containing \textit{cup} and calculated the baseline MCC subject by subject as the pre-adjusted sensitivity measure. Since each subject viewed distinct images during training, we accounted for distribution shifts in both the size and frequency of cup appearances across subjects. To ensure a fair comparison, we adjusted the MCC by normalizing it with the subject-wise average pixel ratio. The resulting weighted MCC is visualized in Figure \ref{fig:cup_mcc}.

\subsection{More Analytical and Visual Results}

\subsubsection{Voxel Sensitivity} 
To examine voxel sensitivities, we analyzed the mean (x-axis) and standard deviation (y-axis) of voxel activations across five brain ROIS for objects \textit{bench} and \textit{chair}, as presented in Figure \ref{fig:vo_mean_std}. The results indicate a quadratic relationship in voxel sensitivity across these regions, allowing us to classify voxels into three distinct groups based on their mean activation and variability (standard deviation). Each group reflects a unique role in semantic decoding within visual regions:
 \begin{itemize}[leftmargin=*]
     \item Bystanders – This group, characterized by the lowest mean and standard deviation, consists of voxels that consistently contribute minimal information to the semantic decoding of visual stimuli. These voxels are either not responsible for distinguishing specific objects (\textit{bench} and \textit{chair}) or likely located in regions less involved in object discrimination, and instead providing generalized but stable responses across diverse stimuli.
     \item Discriminators – This higher mean and the highest standard deviation group includes voxels that show selective, highly variable responses, playing a key role in differentiating between features and supporting object-specific sensitivity. These voxels likely drive the flexibility needed for nuanced and accurate decoding of semantic information in visual stimuli.
     \item Supporters – The highest mean, low standard deviation voxels, characterized by strong, consistent activation, likely represent core object features and provide a stable foundation for robust and invariant decoding across subjects. They likely provide stable, foundational support for correctly classifying objects across different conditions.
 \end{itemize}
These findings suggest that voxel sensitivity patterns vary across the visual hierarchy, with each group contributing distinct vision information-processing roles in object recognition in the brain.

\begin{figure}[h]
  \centering
  \includegraphics[width=1\textwidth]{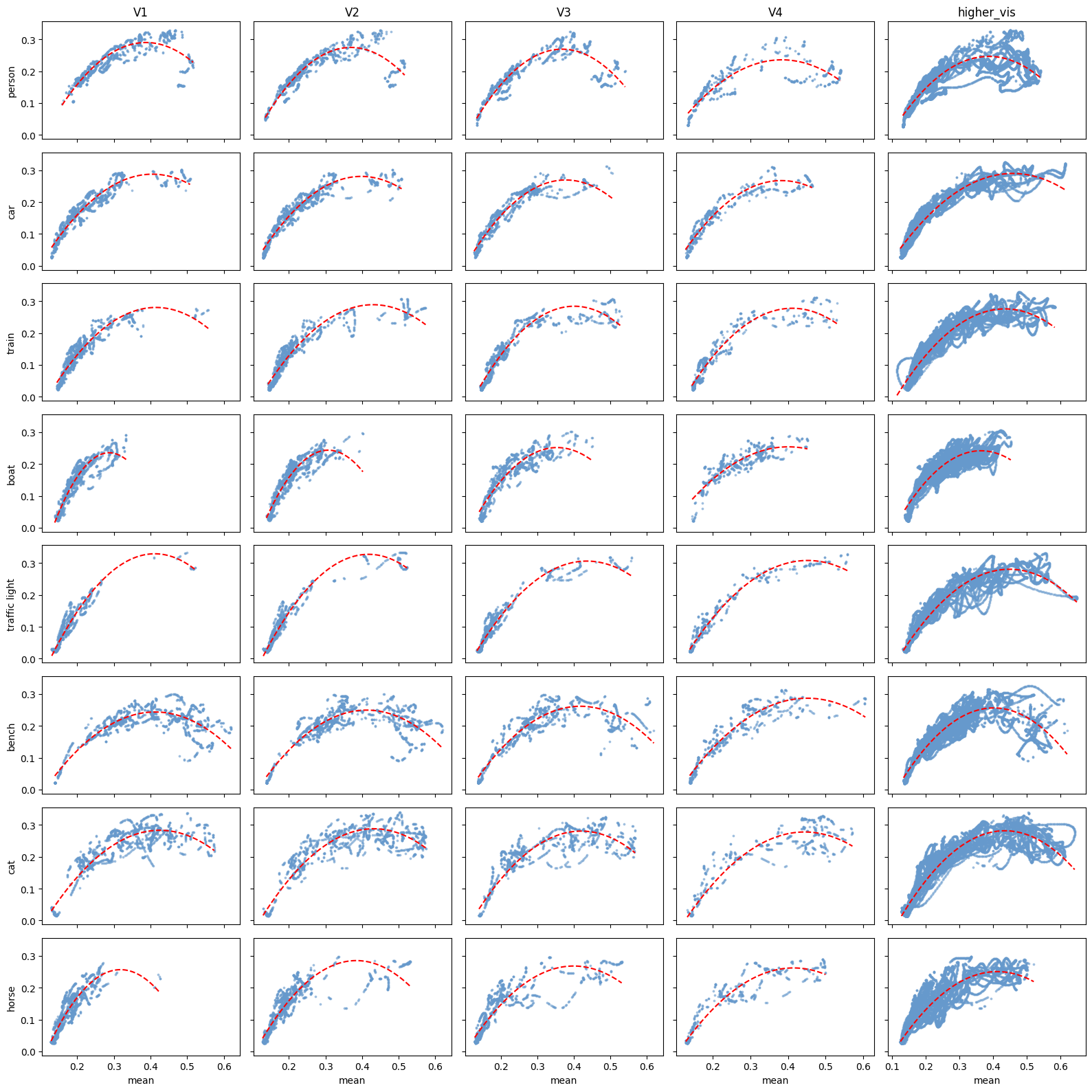}
  \caption{Object-Voxel activations (std v.s. mean) by vision ROIs for subj01}
\end{figure}\label{fig:vo_mean_std}

\subsubsection{Single-subject 1D Activation Pattern}
Similar to Figure \ref{fig:vo_median} in the main paper, we provide more visualization results on 1D Object-Voxel activation below:

\begin{figure*}[ht]
  \centering
  \includegraphics[width=.96\textwidth]{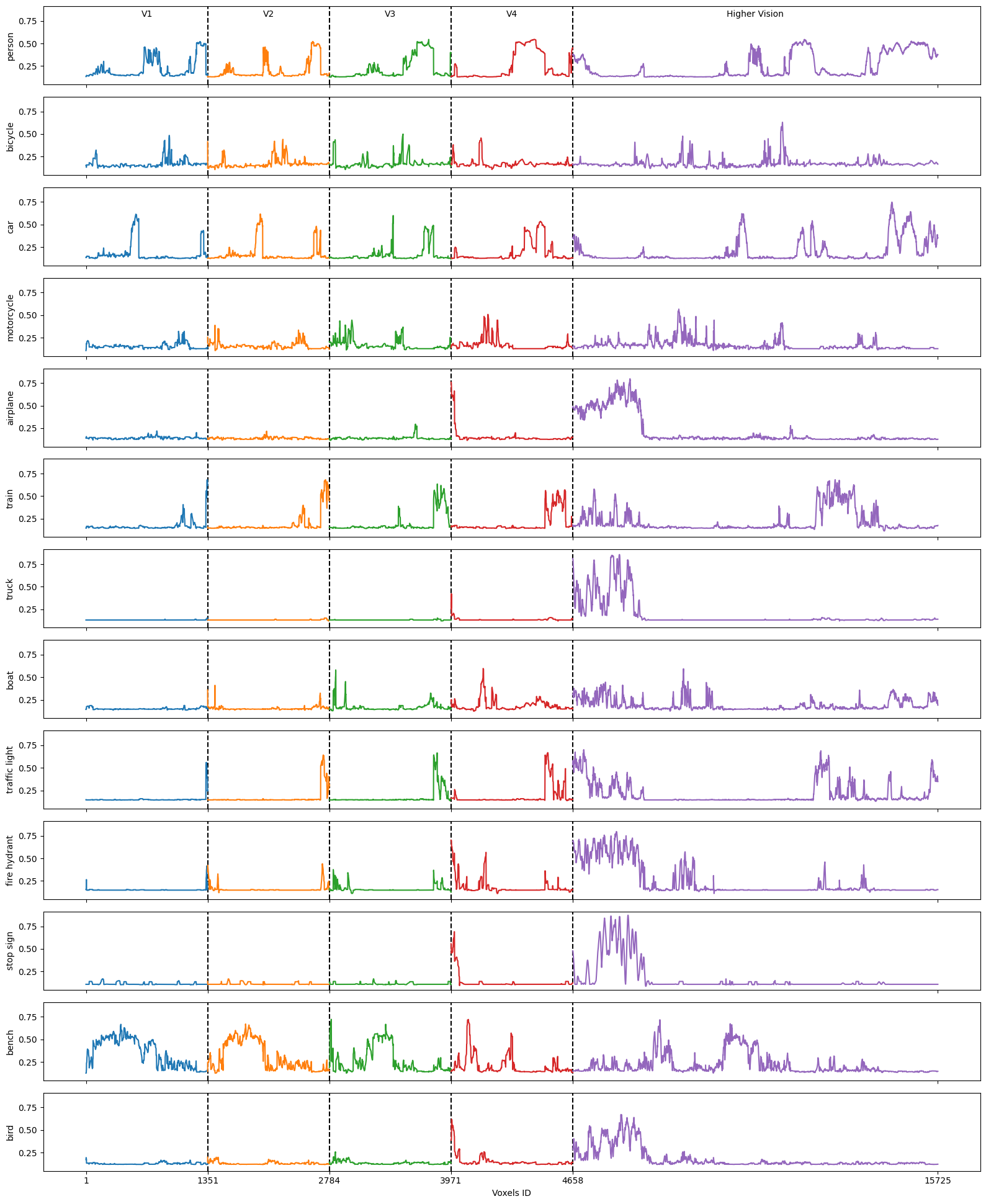}
  \caption{1D Object-Voxel activations by brain vision ROIs for subj01.}
\end{figure*}

\begin{figure*}[ht]
  \centering
  \includegraphics[width=1\textwidth]{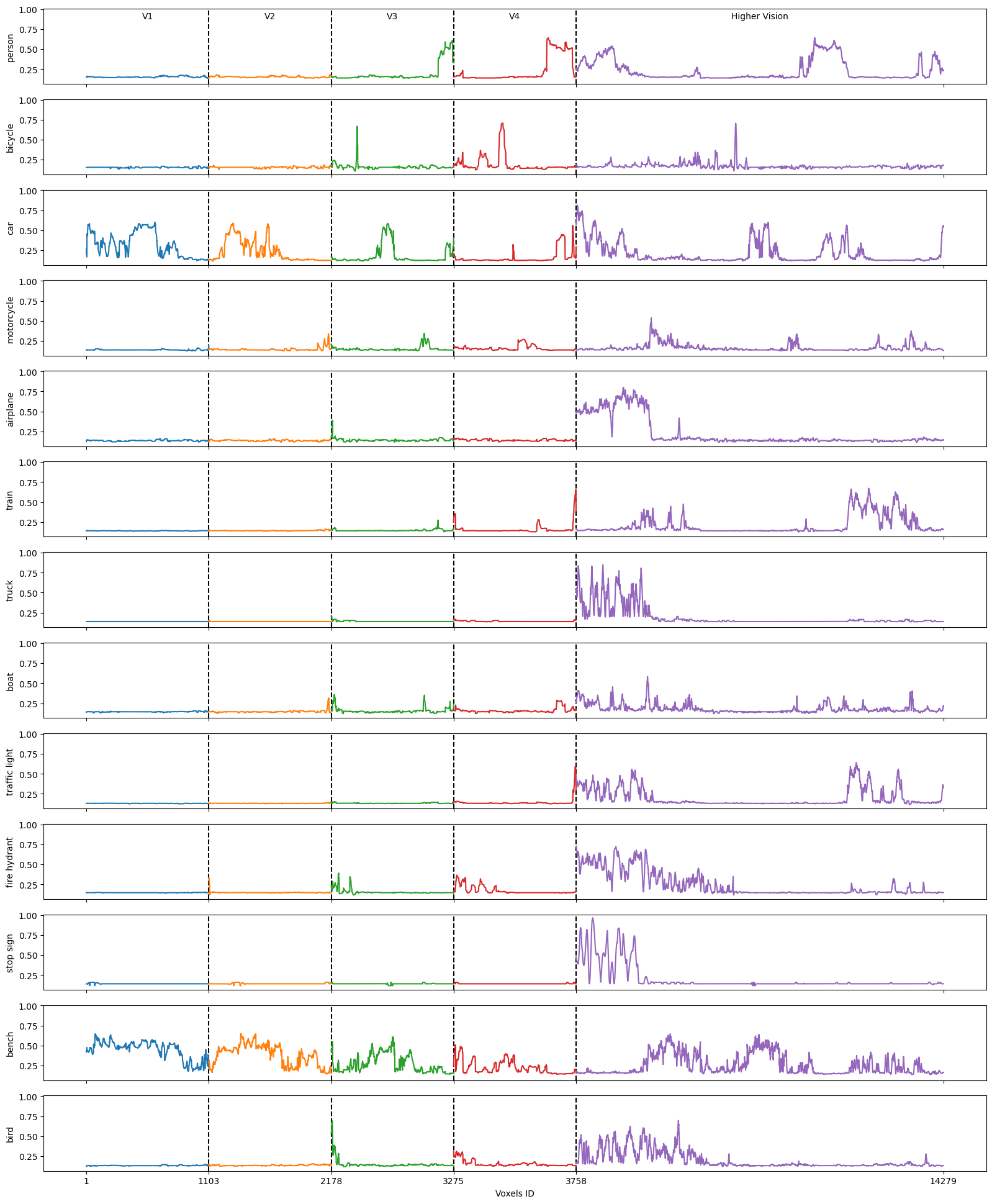}
  \caption{1D Object-Voxel activations by brain vision ROIs for subj02.}
\end{figure*}

\begin{figure*}[ht]
  \centering
  \includegraphics[width=1\textwidth]{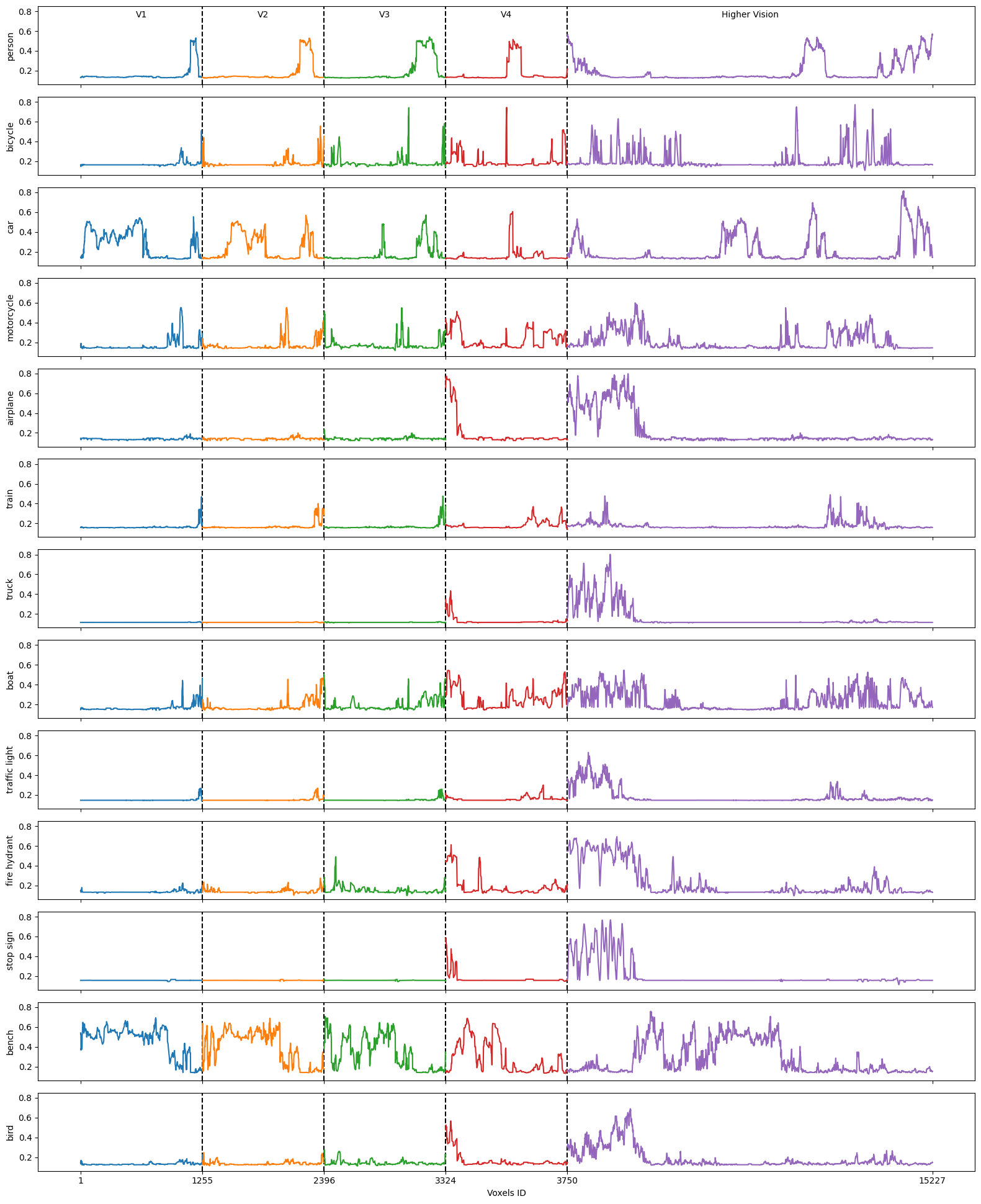}
  \caption{1D Object-Voxel activations by brain vision ROIs for subj03.}
\end{figure*}

\begin{figure*}[ht]
  \centering
  \includegraphics[width=1\textwidth]{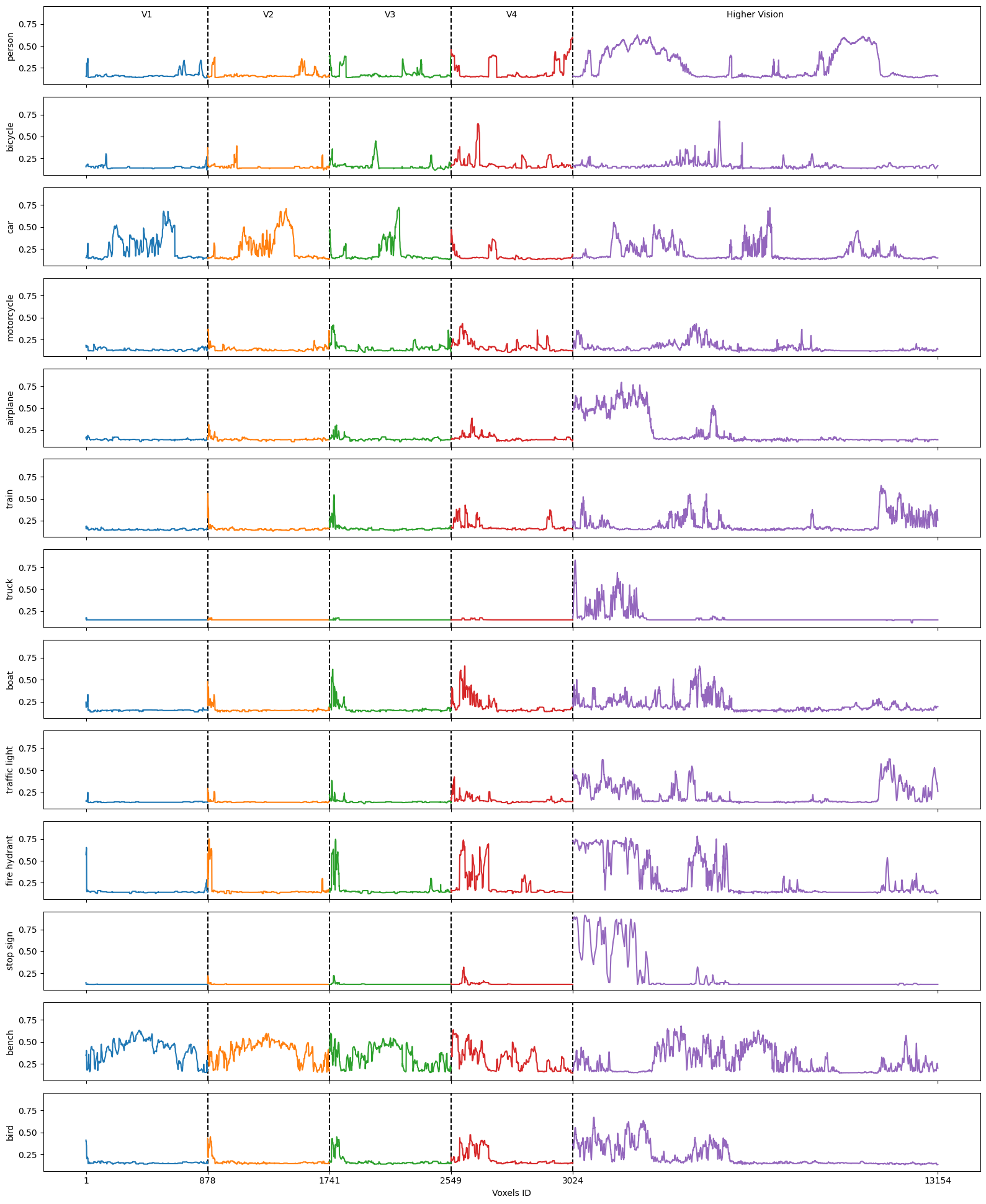}
  \caption{1D Object-Voxel activations by brain vision ROIs for subj04.}
\end{figure*}

\begin{figure*}[ht]
  \centering
  \includegraphics[width=1\textwidth]{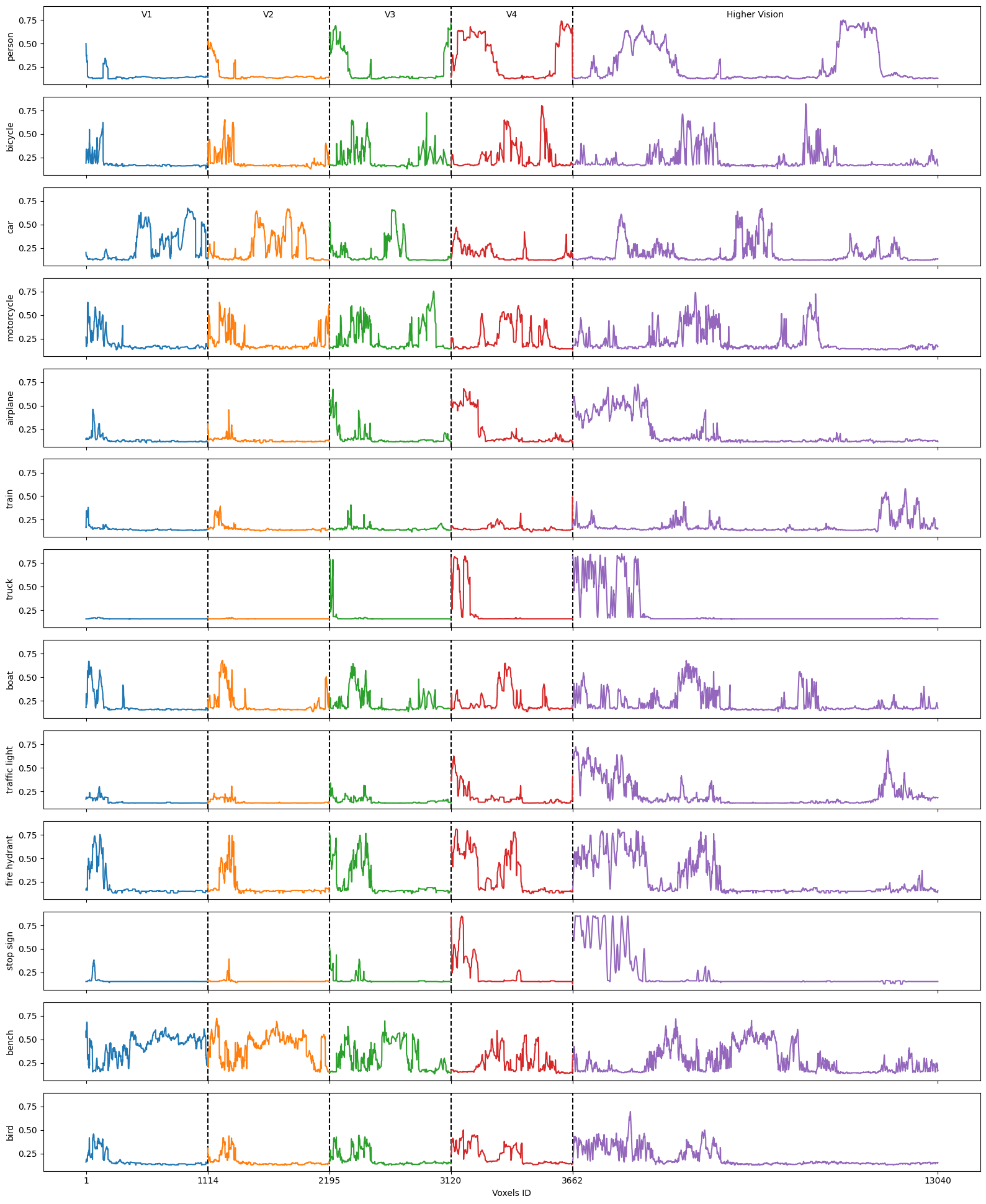}
  \caption{1D Object-Voxel activations by brain vision ROIs for subj05.}
\end{figure*}

\begin{figure*}[ht]
  \centering
  \includegraphics[width=1\textwidth]{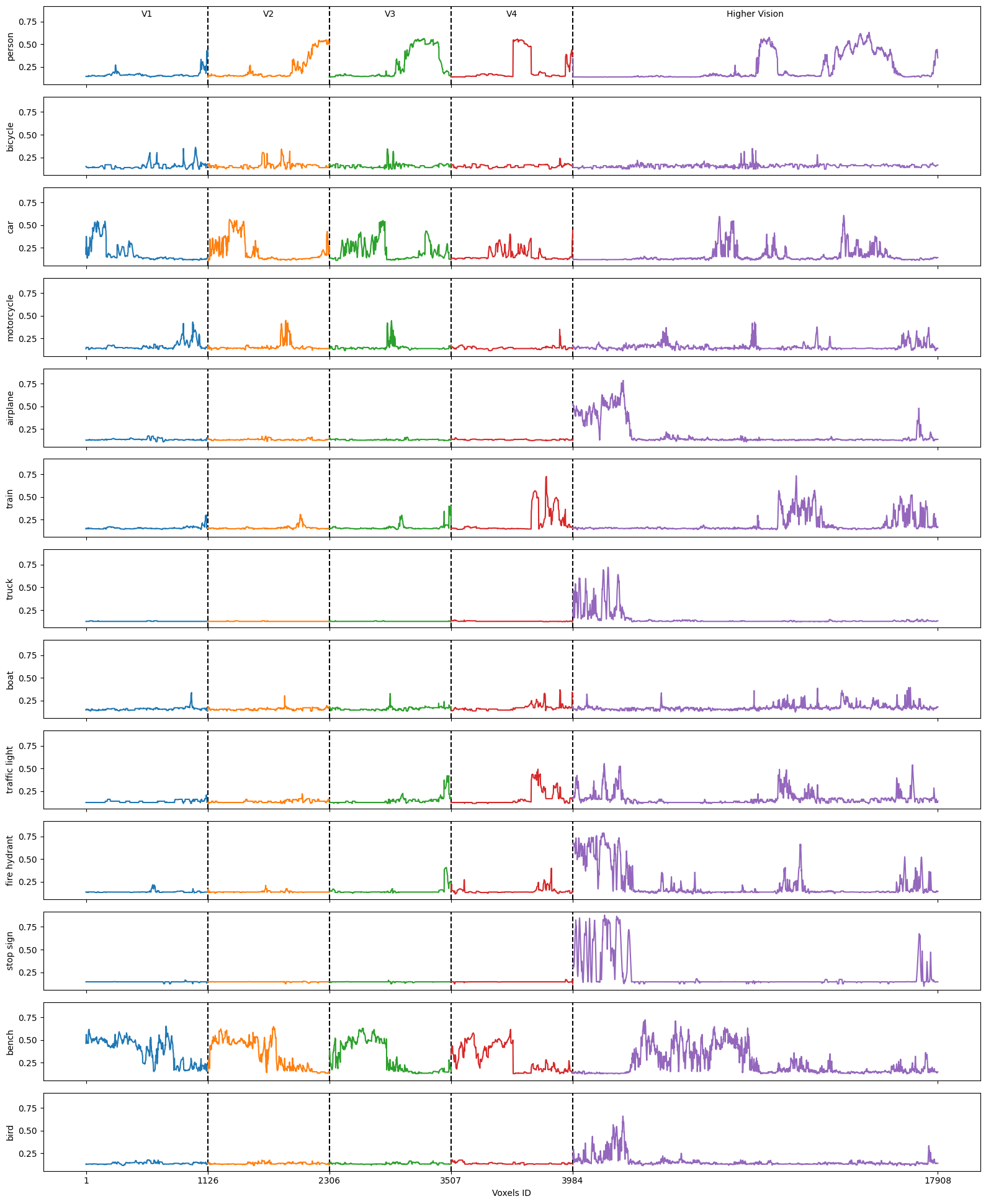}
  \caption{1D Object-Voxel activations by brain vision ROIs for subj06.}
\end{figure*}

\begin{figure*}[ht]
  \centering
  \includegraphics[width=1\textwidth]{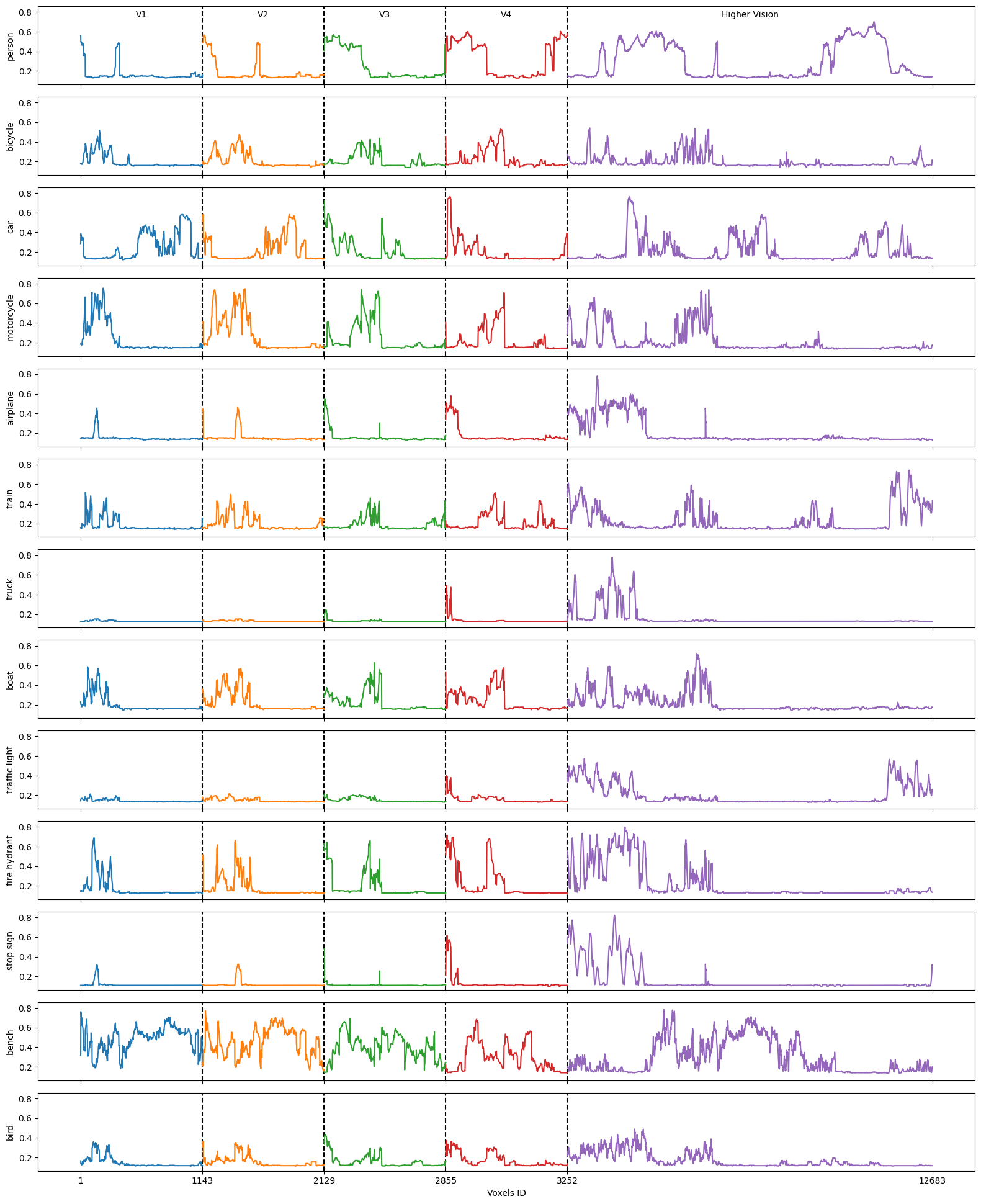}
  \caption{1D Object-Voxel activations by brain vision ROIs for subj07.}
\end{figure*}

\begin{figure*}[ht]
  \centering
  \includegraphics[width=1\textwidth]{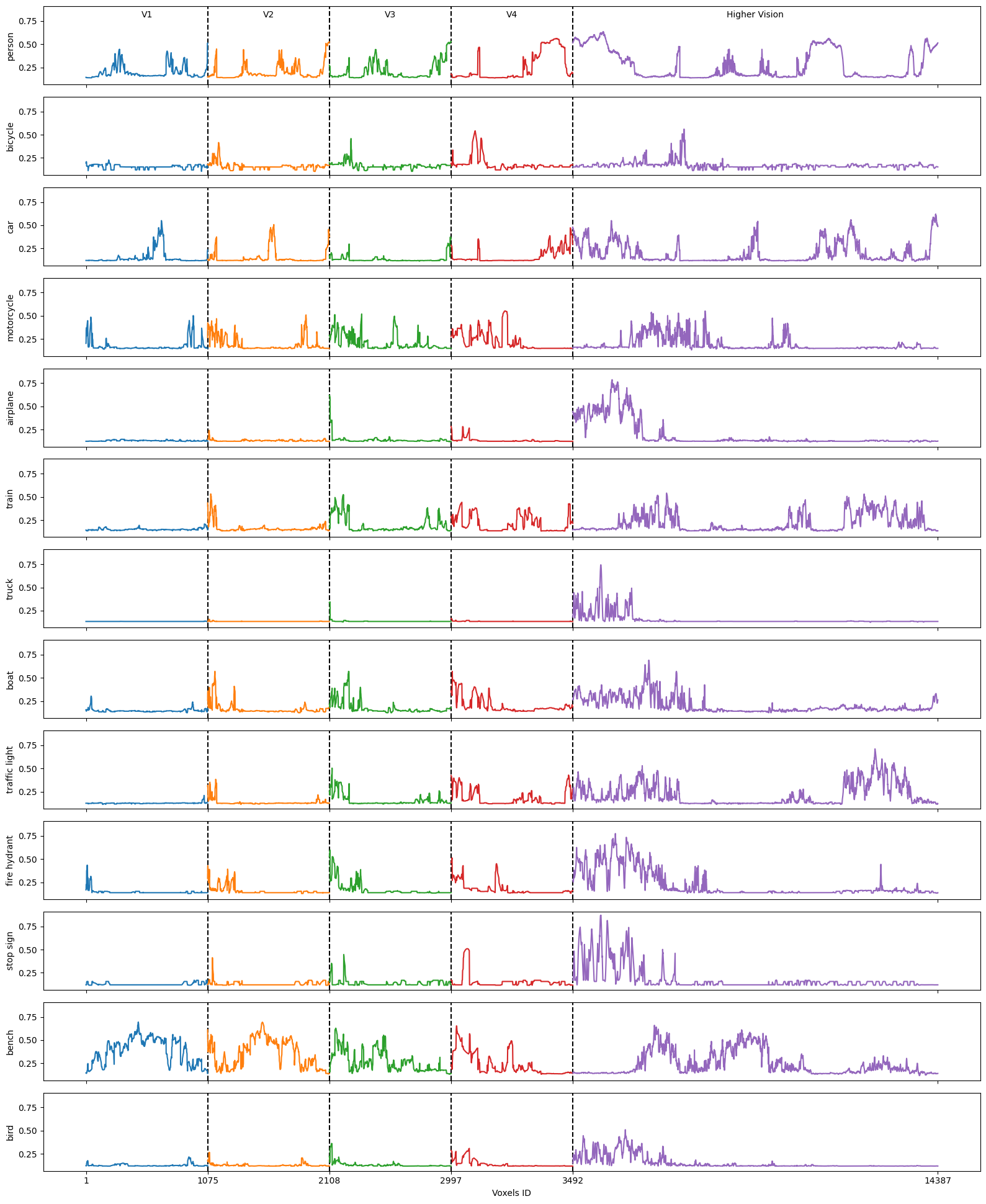}
  \caption{1D Object-Voxel activations by brain vision ROIs for subj08.}
\end{figure*}
\clearpage

\subsubsection{Cross-subject 3D Activation Pattern}
Similar to Figure \ref{fig:vo_3d_mean} in the main paper, we provide more visualization results on 3D Object-Voxel activation below:

\begin{figure*}[ht]
  \centering
  \includegraphics[width=1\textwidth]{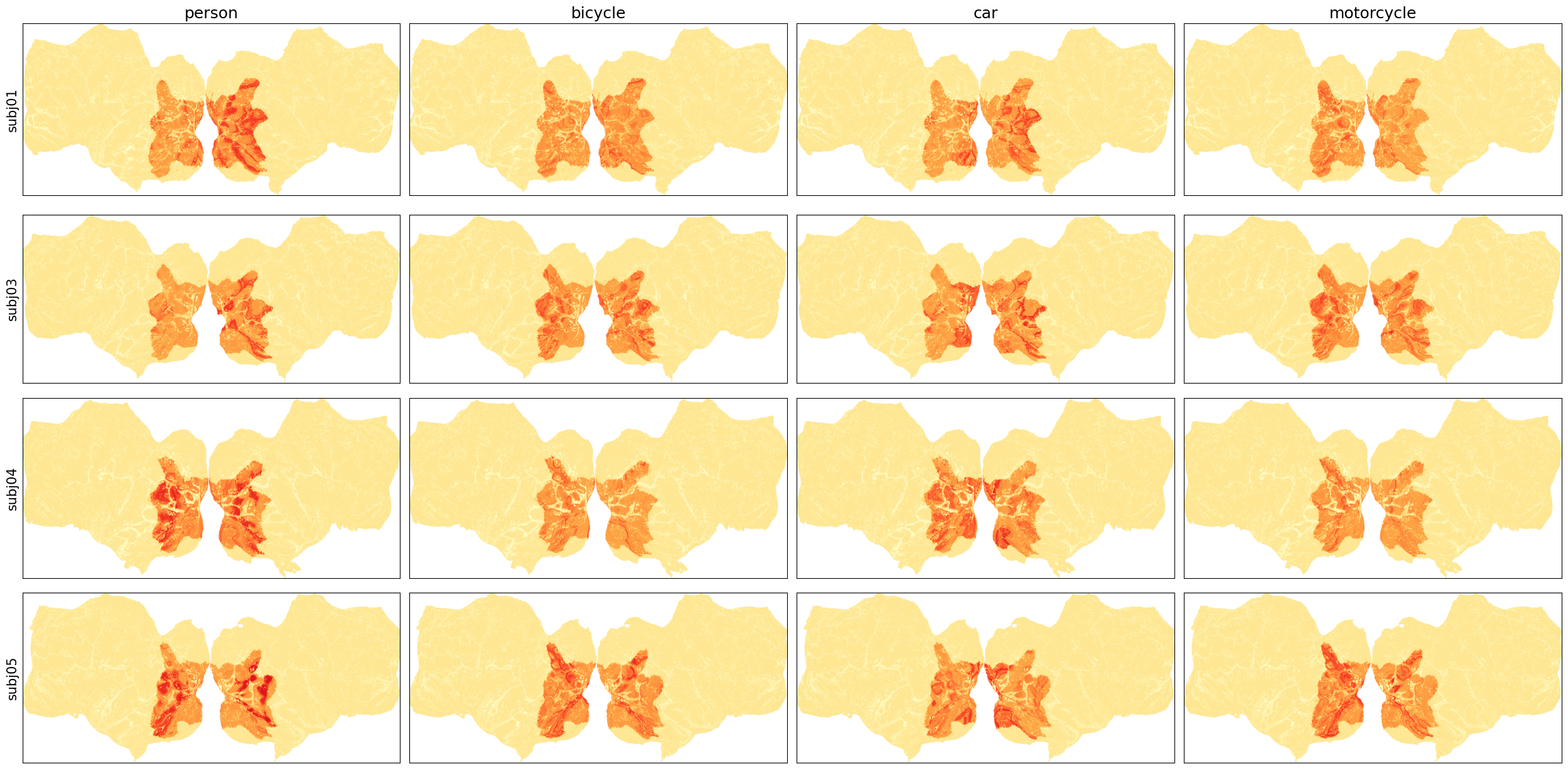}
  \caption{3D Object-Voxel activations of \textit{person}, \textit{bicycle}, \textit{car}, and \textit{motorcycle} for subj01, subj03, subj04, and subj05.}
\end{figure*}
\begin{figure*}[ht]
  \centering
  \includegraphics[width=1\textwidth]{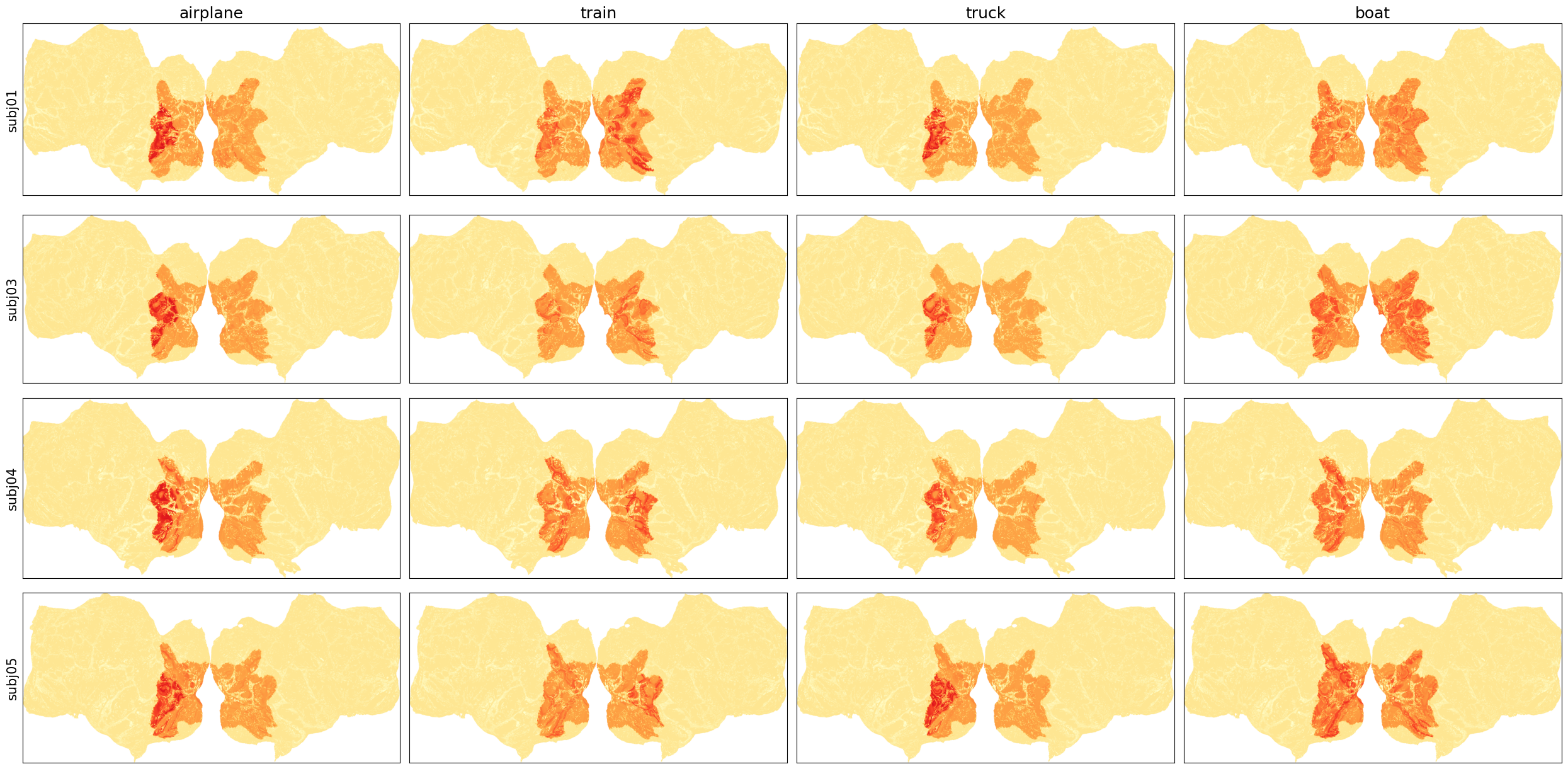}
  \caption{3D Object-Voxel activations of \textit{airplane}, \textit{train}, \textit{truck}, and \textit{boat} for subj01, subj03, subj04, and subj05.}
\end{figure*}
\begin{figure*}[ht]
  \centering
  \includegraphics[width=1\textwidth]{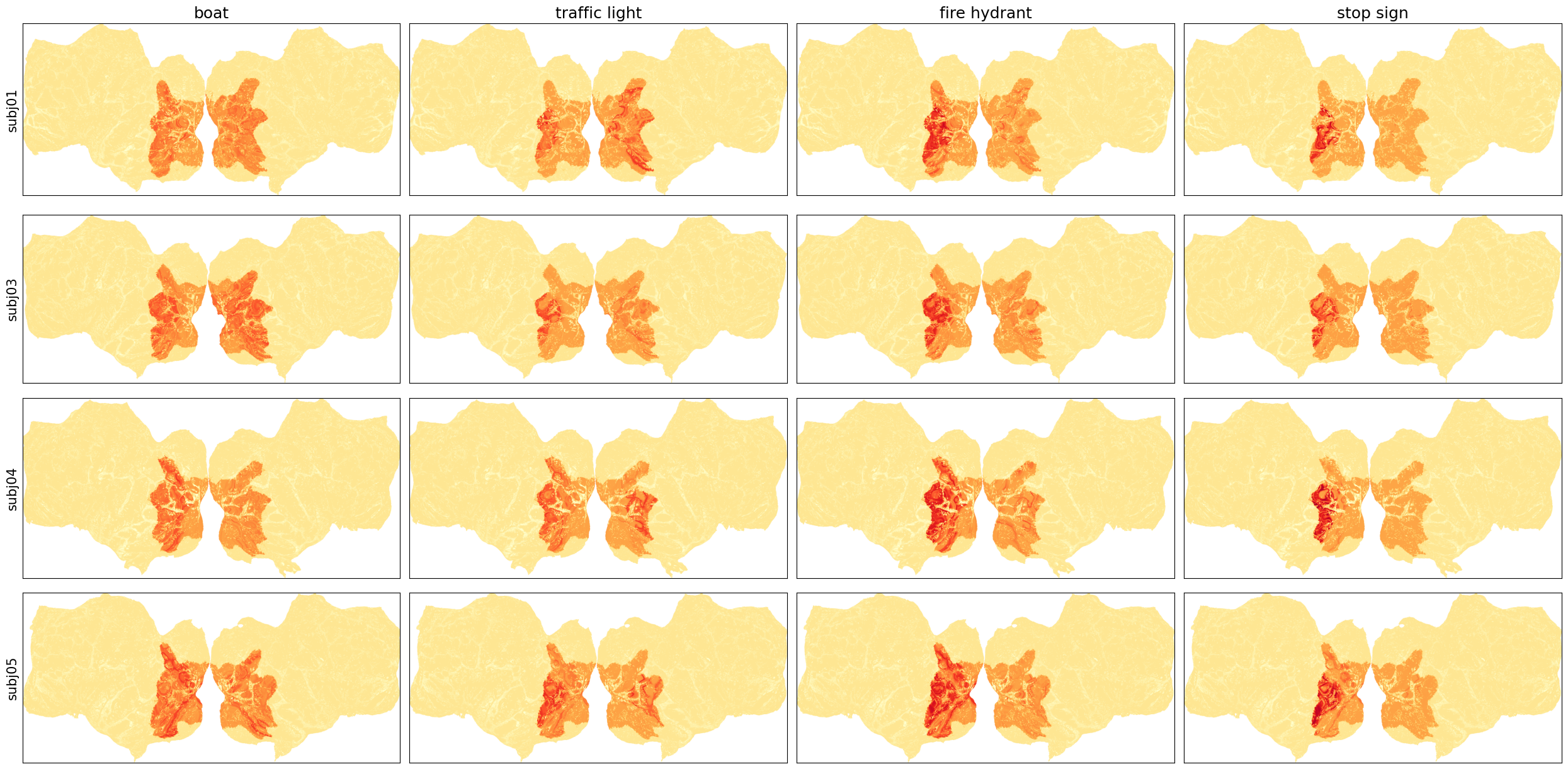}
    \caption{3D Object-Voxel activations of \textit{boat}, \textit{traffic light}, \textit{fire hydrant}, and \textit{stop sign} for subj01, subj03, subj04, and subj05.}
\end{figure*}
\begin{figure*}[ht]
  \centering
  \includegraphics[width=1\textwidth]{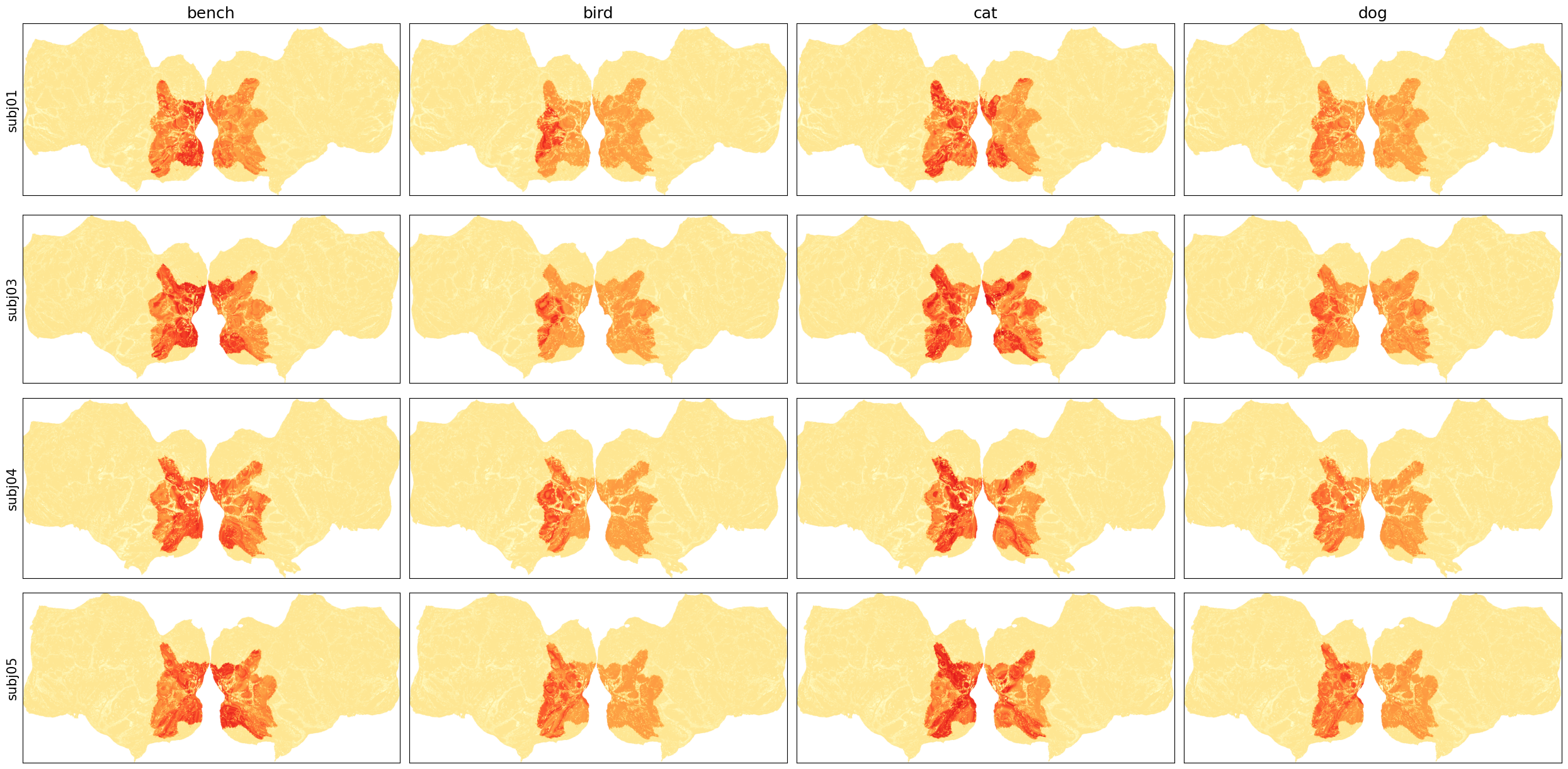}
  \caption{3D Object-Voxel activations of \textit{bench}, \textit{bird}, \textit{cat}, and \textit{dog} for subj01, subj03, subj04, and subj05.}
\end{figure*}
\clearpage

\subsubsection{Variations in Subject Attention}
Similar to Figure 4 in the main paper, we provide more visualization results on variations in subjects' attention to different objects in the same image. The leftmost image shows the visual stimulus. Plots in the second column represent the shared attention across all subjects, and the remaining eight columns show the residual, subject-specific attention alongside predicted probabilities to compare recognition confidence and priority.

\begin{figure*}[t]
  \centering
  \includegraphics[width=1\textwidth]{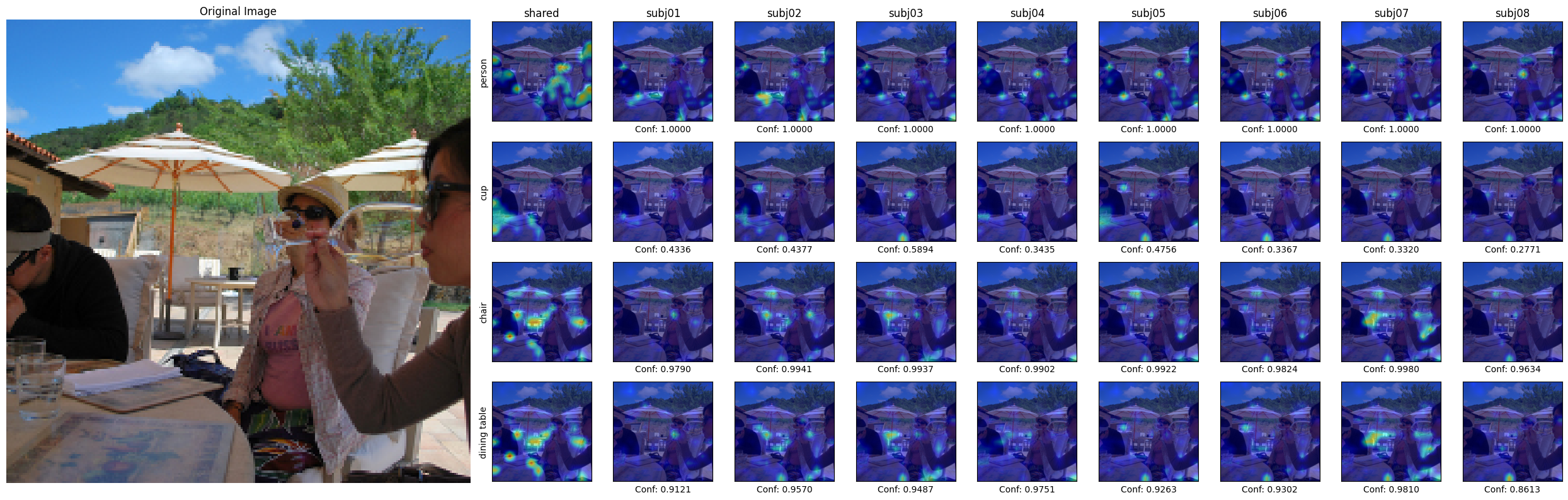}
  \includegraphics[width=1\textwidth]{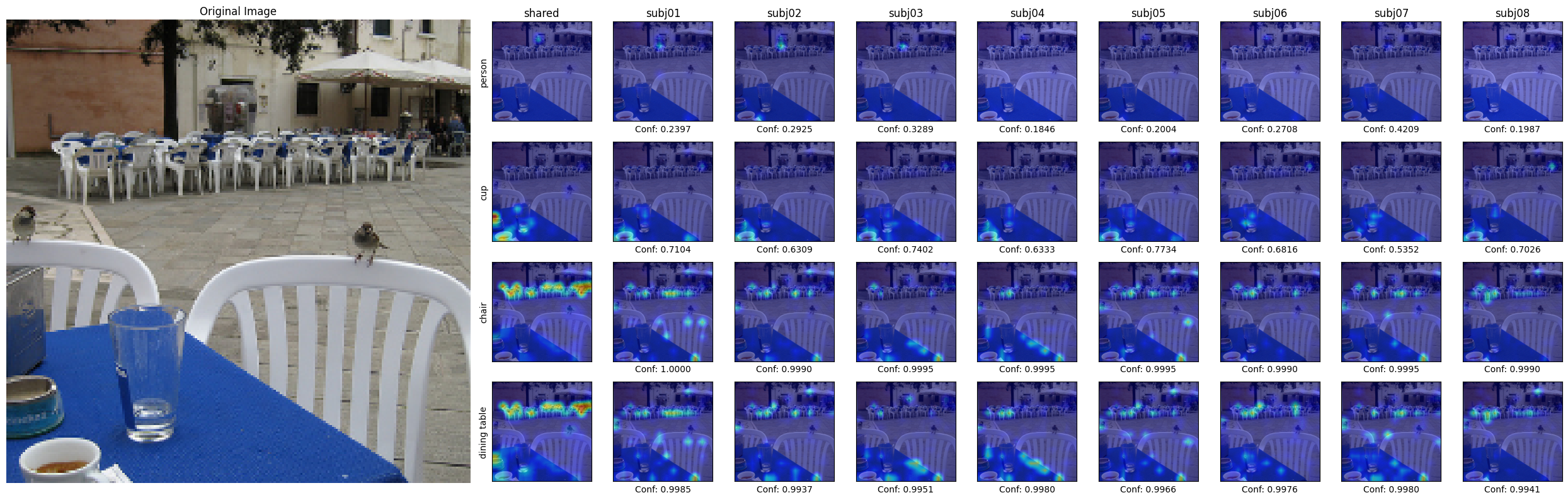}
  \includegraphics[width=1\textwidth]{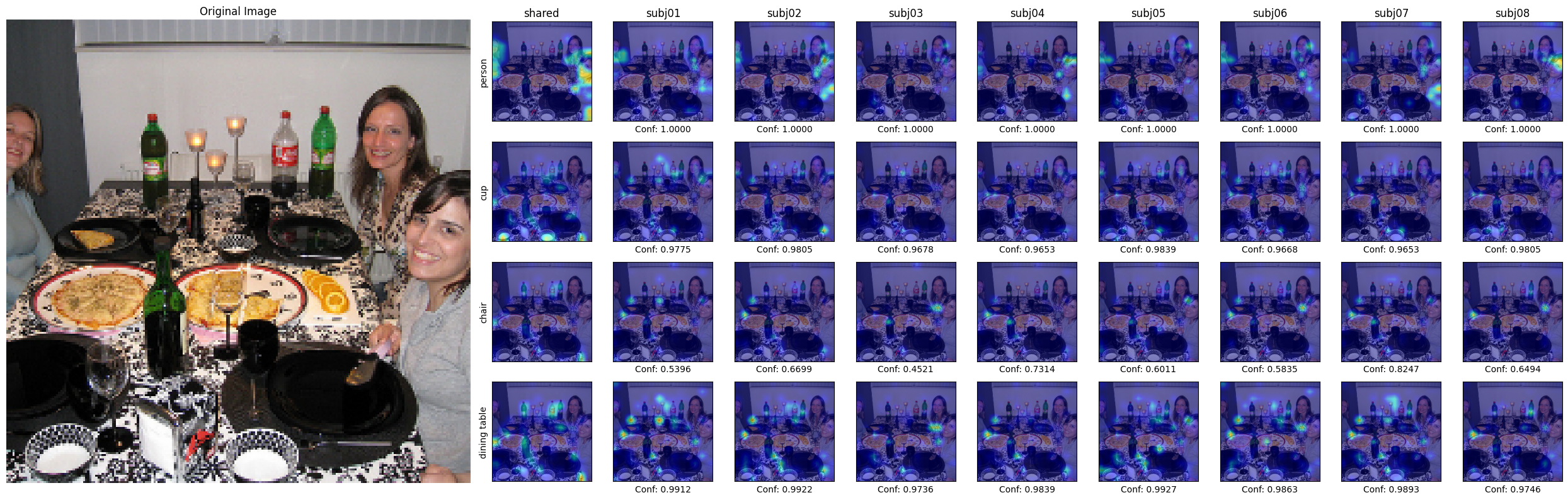}
  \caption{Variations in subjects' attention to different objects. Four objects: \textit{person}, \textit{cup}, \textit{chair}, and \textit{dining table} are selected for visualization.
 }
\end{figure*}

\begin{figure*}[t]
  \centering
  \includegraphics[width=1\textwidth]{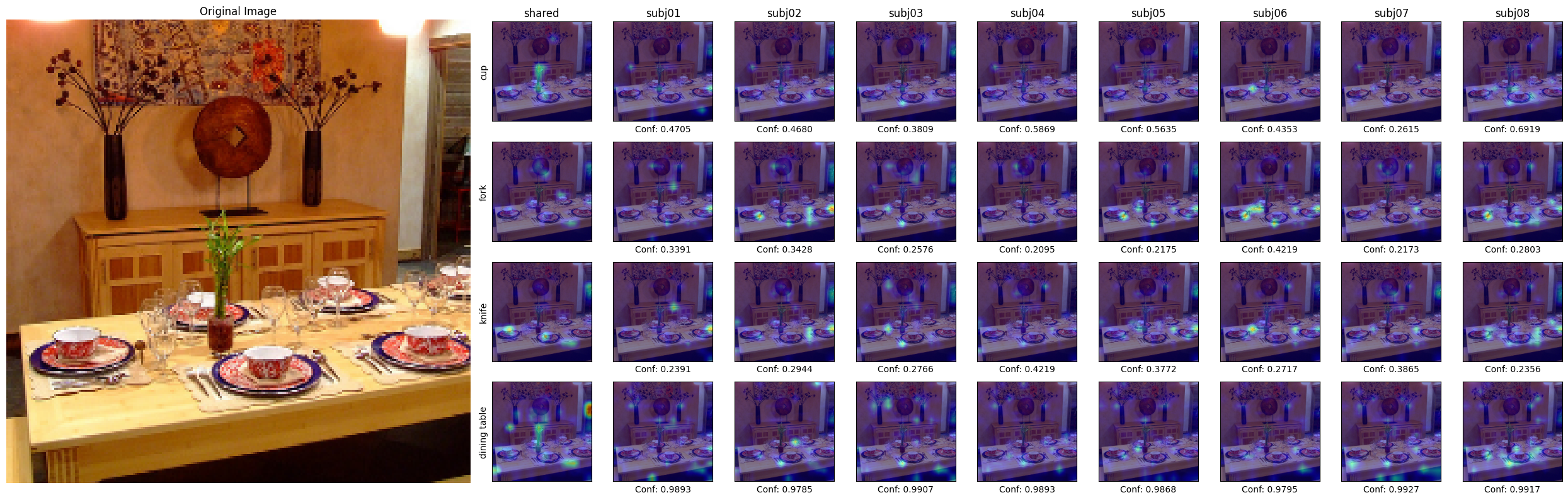}
  \includegraphics[width=1\textwidth]{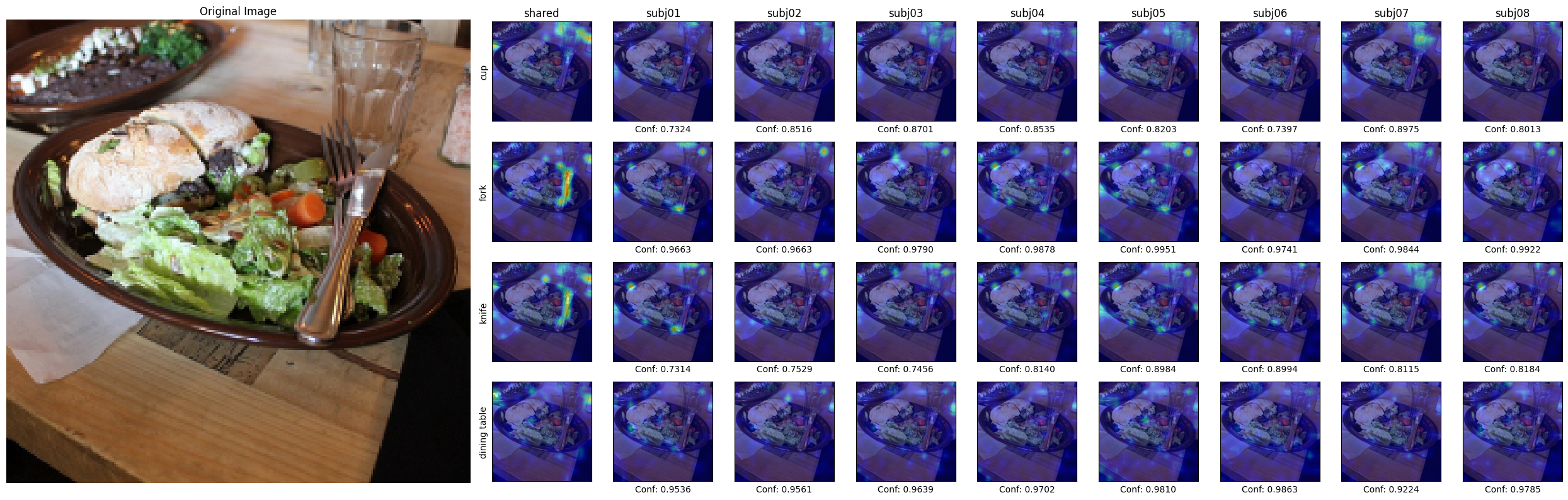}
  \includegraphics[width=1\textwidth]{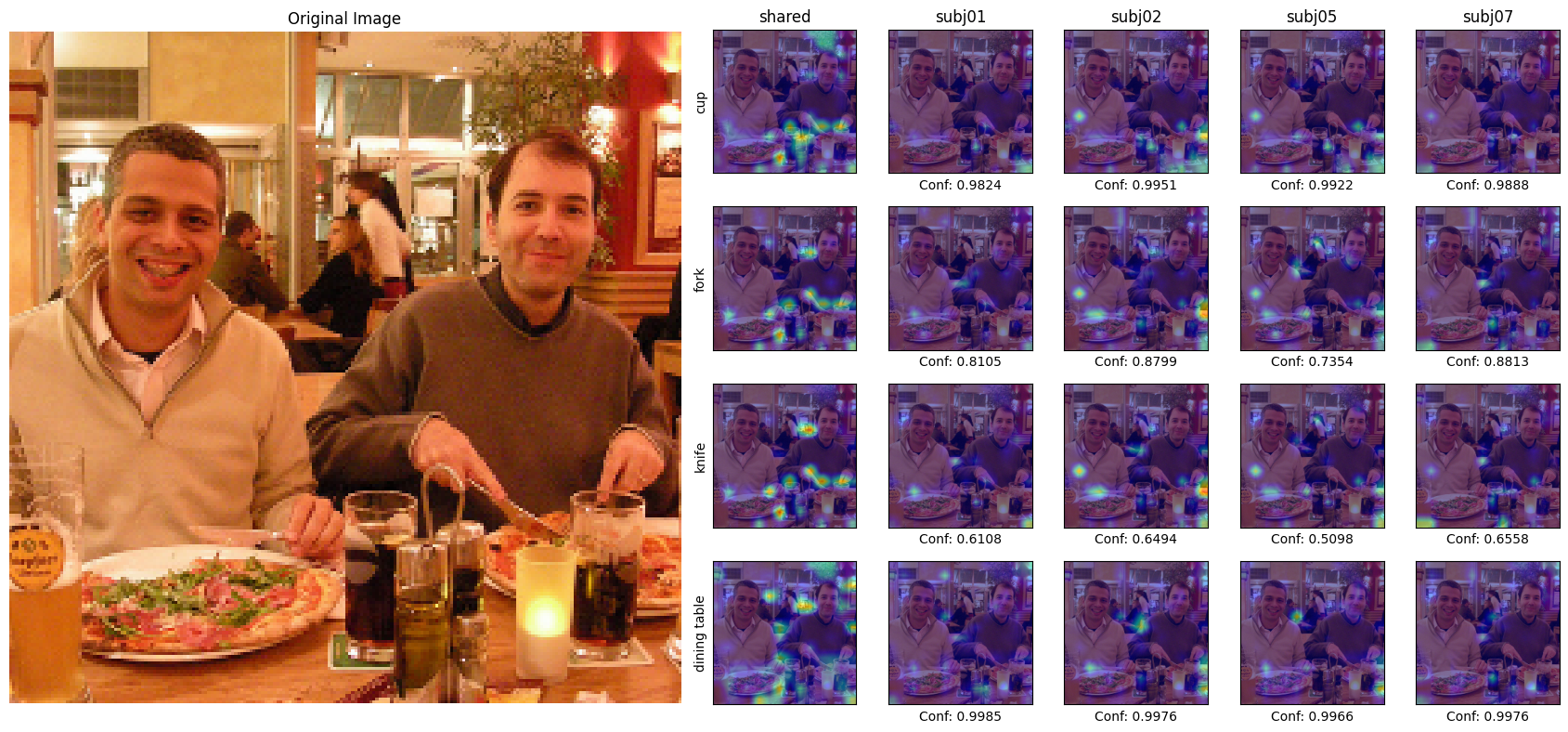}
  \caption{Variations in subjects' attention to different objects. Four objects: \textit{cup}, \textit{fork}, \textit{knife}, and \textit{dining table} are selected for visualization.
 }
\end{figure*}

\begin{figure*}[t]
  \centering
  \includegraphics[width=.95\textwidth]{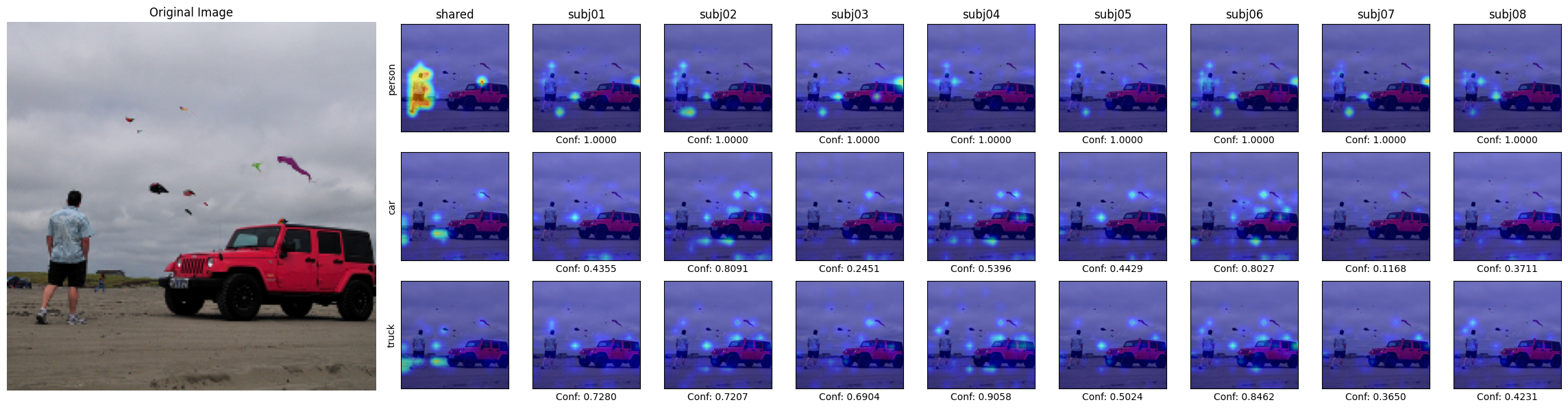}
  \includegraphics[width=.95\textwidth]{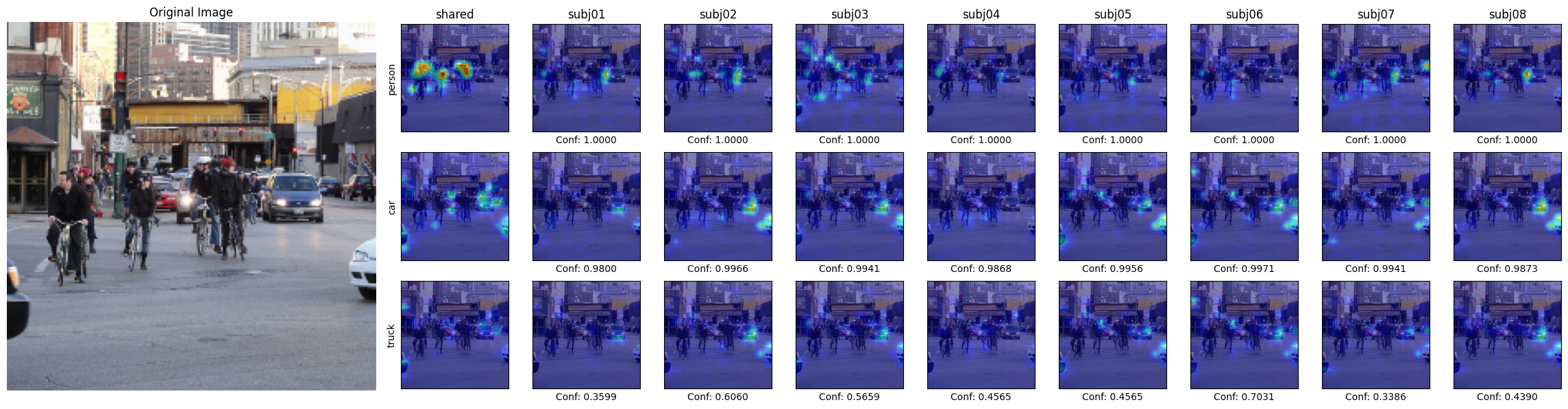}
  \includegraphics[width=.95\textwidth]{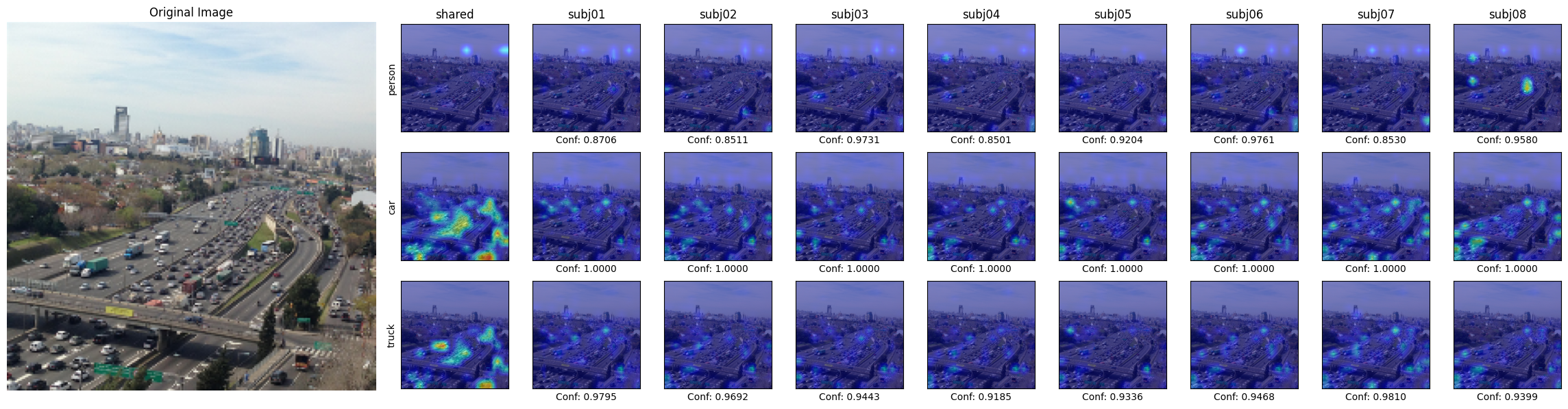}
  \includegraphics[width=.95\textwidth]{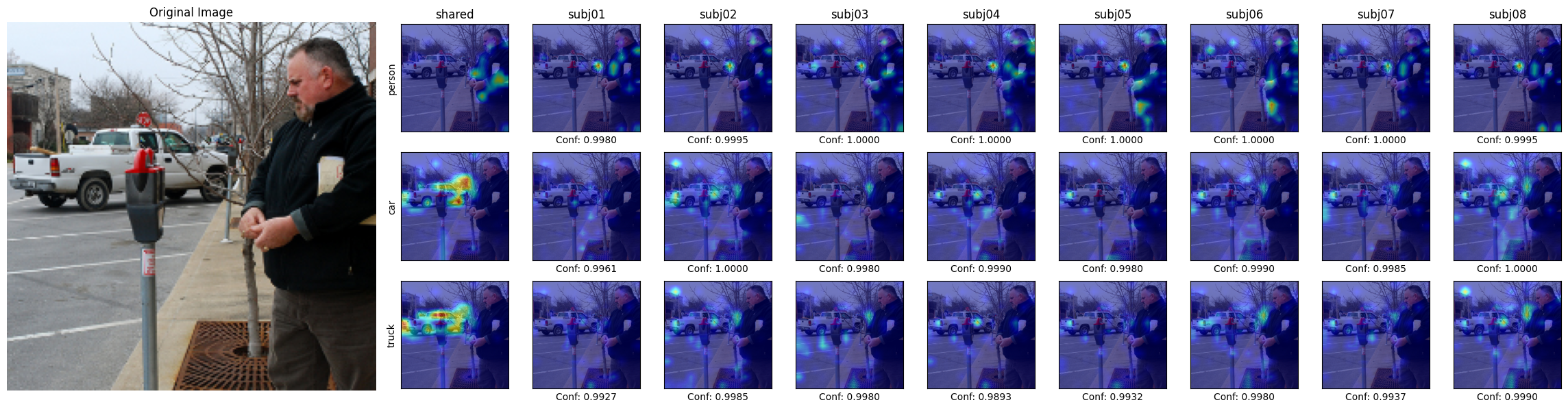}
  \includegraphics[width=.95\textwidth]{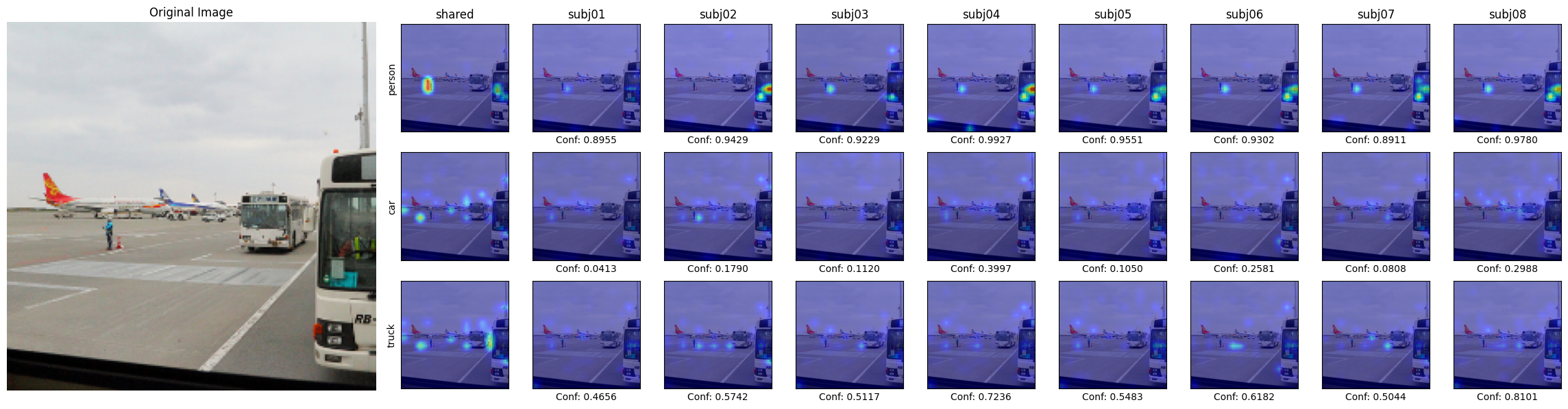}
  \caption{Variations in subjects' attention to different objects. Three objects: \textit{person}, \textit{car}, and \textit{truck} are selected for visualization.
 }
\end{figure*}

\begin{figure*}[t]
  \centering
  \includegraphics[width=.94\textwidth]{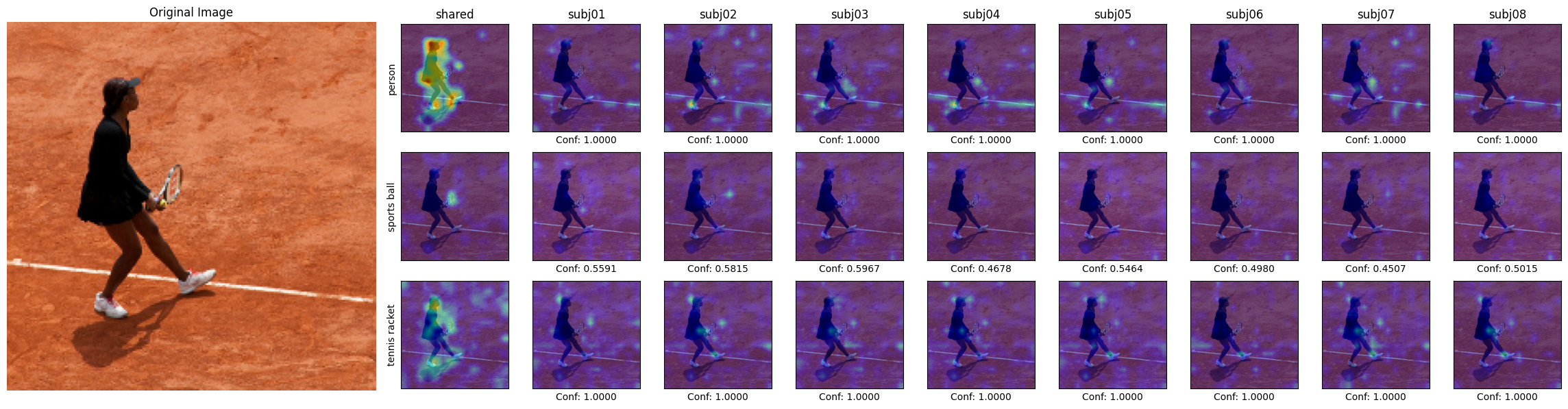}
  \includegraphics[width=.94\textwidth]{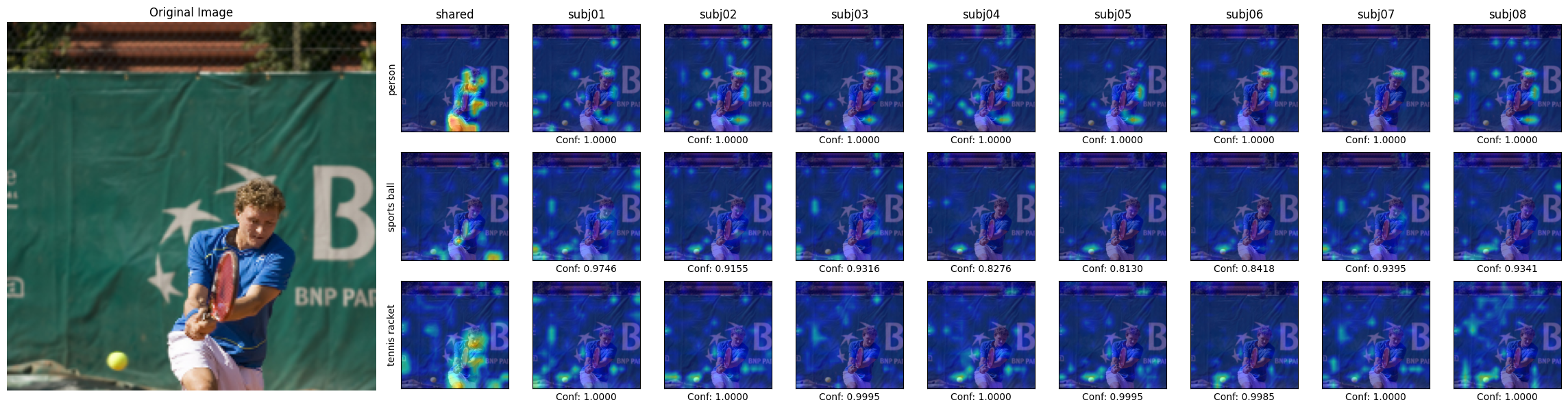}
  \includegraphics[width=.94\textwidth]{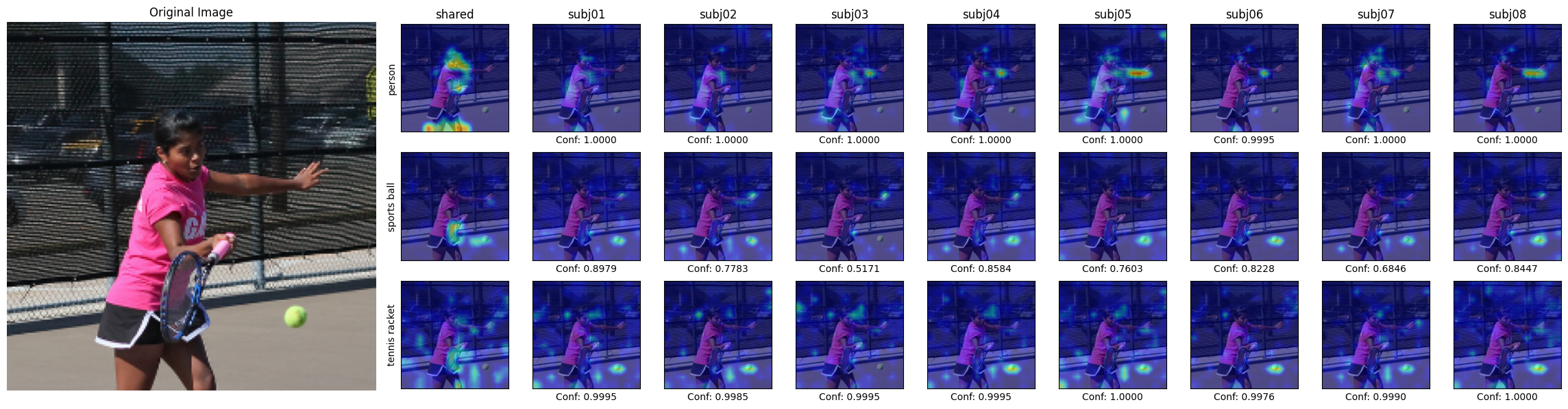}
  \includegraphics[width=.94\textwidth]{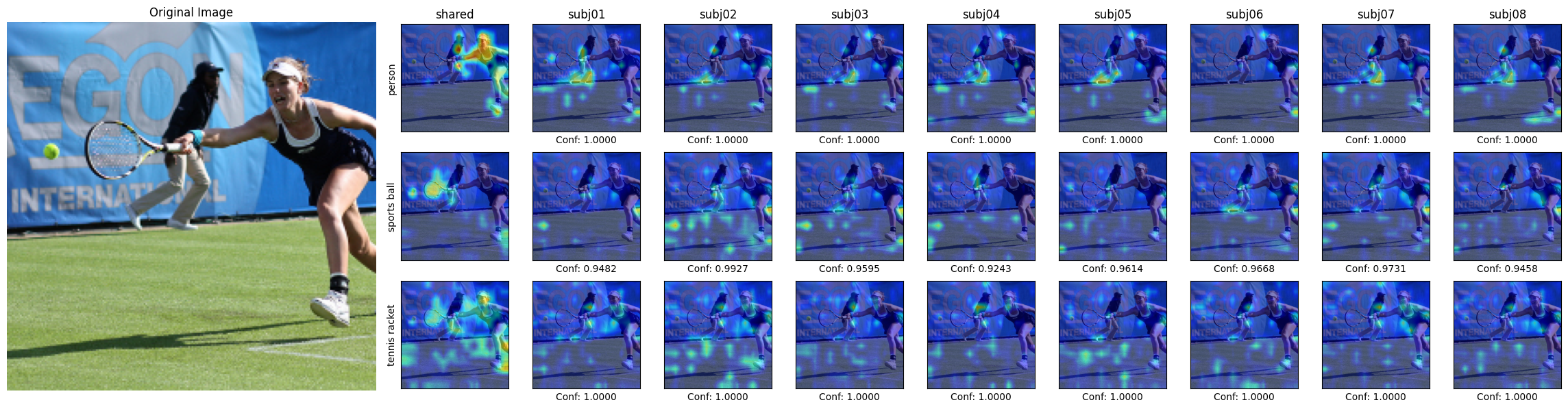}
  \includegraphics[width=.94\textwidth]{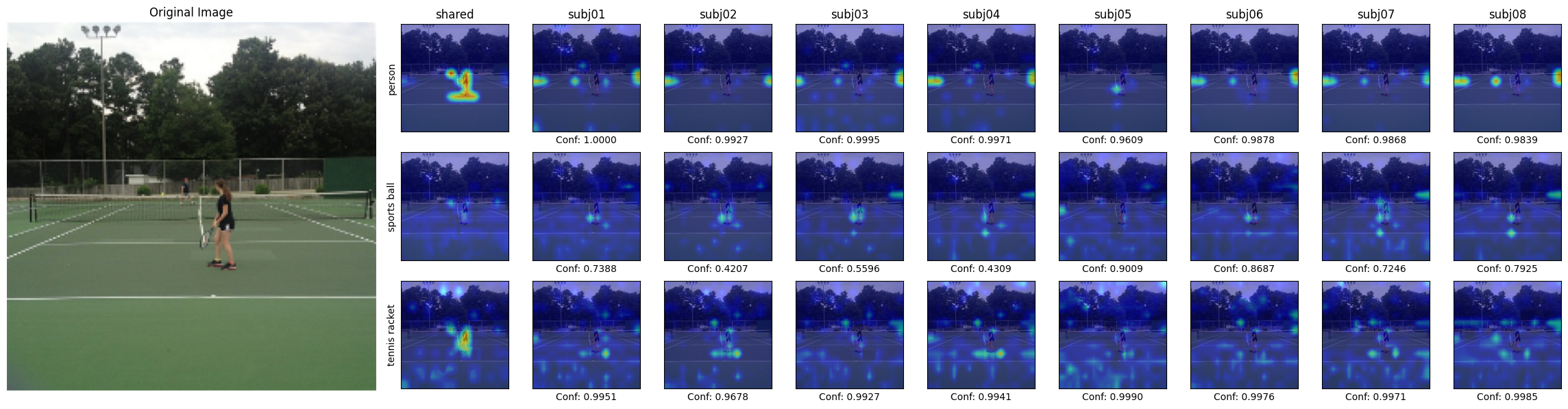}
  \caption{Variations in subjects' attention to different objects. Three objects: \textit{person}, \textit{sports ball}, and \textit{tennis racket} are selected for visualization.
 }
\end{figure*}

\clearpage

%% file: sec/A2_checklist.tex
\section*{NeurIPS Paper Checklist}

\begin{enumerate}

\item {\bf Claims}
    \item[] Question: Do the main claims made in the abstract and introduction accurately reflect the paper's contributions and scope?
    \item[] Answer: \answerYes{} 
    \item[] Justification: we clearly state our contributions that match claims made in the abstract and introduction. we also did theoretic proof and conduct extensively experiments to validate our claims in the result section in the main paper and in Appendix

\item {\bf Limitations}
    \item[] Question: Does the paper discuss the limitations of the work performed by the authors?
    \item[] Answer: \answerYes{} 
    \item[] Justification: Due to page limit, we don't include a limitation section in the main paper. Instead, we point out the limitation of the proposed method in Appendix \ref{subsec:limit}.

\item {\bf Theory assumptions and proofs}
    \item[] Question: For each theoretical result, does the paper provide the full set of assumptions and a complete (and correct) proof?
    \item[] Answer: \answerNA{} 
    \item[] Justification: We don't make any theoretic contributions in the paper, but we do rely on the assumption that the subject and object information is linear entangled in the latent fMRI space.

    \item {\bf Experimental result reproducibility}
    \item[] Question: Does the paper fully disclose all the information needed to reproduce the main experimental results of the paper to the extent that it affects the main claims and/or conclusions of the paper (regardless of whether the code and data are provided or not)?
    \item[] Answer: \answerYes{} 
    \item[] Justification: All details necessary for reproducing the experimental results are fully described in either the main paper or appendix. The code will be released upon acceptance.

\item {\bf Open access to data and code}
    \item[] Question: Does the paper provide open access to the data and code, with sufficient instructions to faithfully reproduce the main experimental results, as described in supplemental material?
    \item[] Answer: \answerYes{} 
    \item[] Justification: In appendix, we describe the used dataset as well as the version of publicly available pretrained deep models we rely on in details.

\item {\bf Experimental setting/details}
    \item[] Question: Does the paper specify all the training and test details (e.g., data splits, hyperparameters, how they were chosen, type of optimizer, etc.) necessary to understand the results?
    \item[] Answer: \answerYes{} 
    \item[] Justification: Please refer to the implementation details in appeix for all the details regarding hyperparameter choice, preprocessing, etc.

\item {\bf Experiment statistical significance}
    \item[] Question: Does the paper report error bars suitably and correctly defined or other appropriate information about the statistical significance of the experiments?
    \item[] Answer: \answerNo{} 
    \item[] Justification: We don't do statistical significance check simply because our model outperforms existing models by a large margin that don't need a statistical test to prove it. Also, the model is trained and tested on a large dataset. It would be time-consuming to do such test.

\item {\bf Experiments compute resources}
    \item[] Question: For each experiment, does the paper provide sufficient information on the computer resources (type of compute workers, memory, time of execution) needed to reproduce the experiments?
    \item[] Answer: \answerYes{} 
    \item[] Justification: We list  the computing resources needed for the proposed method, whose parameters are publicly available.
    
\item {\bf Code of ethics}
    \item[] Question: Does the research conducted in the paper conform, in every respect, with the NeurIPS Code of Ethics \url{https://neurips.cc/public/EthicsGuidelines}?
    \item[] Answer: \answerYes{} 
    \item[] Justification: We closely follow the code of ethics by NeurIPS.

\item {\bf Broader impacts}
    \item[] Question: Does the paper discuss both potential positive societal impacts and negative societal impacts of the work performed?
    \item[] Answer: \answerNA{} 
    \item[] Justification: The paper is an exploring work and far away from application. So, it may not have any societal impacts for now.
    
\item {\bf Safeguards}
    \item[] Question: Does the paper describe safeguards that have been put in place for responsible release of data or models that have a high risk for misuse (e.g., pretrained language models, image generators, or scraped datasets)?
    \item[] Answer: \answerNo{} 
    \item[] Justification: we don't think the rease of our data or models will have a high risk for misuse.

\item {\bf Licenses for existing assets}
    \item[] Question: Are the creators or original owners of assets (e.g., code, data, models), used in the paper, properly credited and are the license and terms of use explicitly mentioned and properly respected?
    \item[] Answer: \answerYes{} 
    \item[] Justification: The data and the pretrained deep models are fully described. We give credits to the original creator in the main paper.

\item {\bf New assets}
    \item[] Question: Are new assets introduced in the paper well documented and is the documentation provided alongside the assets?
    \item[] Answer: \answerYes{} 
    \item[] Justification: We will release our code upon acceptance, which include new assets such as trained model weights.

\item {\bf Crowdsourcing and research with human subjects}
    \item[] Question: For crowdsourcing experiments and research with human subjects, does the paper include the full text of instructions given to participants and screenshots, if applicable, as well as details about compensation (if any)? 
    \item[] Answer: \answerNo{} 
    \item[] Justification: Our research relies on human neural signals data, but it has no potential risks for incurred participants.

\item {\bf Institutional review board (IRB) approvals or equivalent for research with human subjects}
    \item[] Question: Does the paper describe potential risks incurred by study participants, whether such risks were disclosed to the subjects, and whether Institutional Review Board (IRB) approvals (or an equivalent approval/review based on the requirements of your country or institution) were obtained?
    \item[] Answer: \answerNo{} 
    \item[] Justification: Our research relies on human neural signals data, but it has no potential risks for incurred participants.

\item {\bf Declaration of LLM usage}
    \item[] Question: Does the paper describe the usage of LLMs if it is an important, original, or non-standard component of the core methods in this research? Note that if the LLM is used only for writing, editing, or formatting purposes and does not impact the core methodology, scientific rigorousness, or originality of the research, declaration is not required.
    \item[] Answer: \answerNA{} 
    \item[] Justification: the core method development in this research does not involve LLMs as any important, original, or non-standard components.

\end{enumerate}

%% file: main.bbl
\begin{thebibliography}{10}

\bibitem{attn_rollout}
Samira Abnar and Willem Zuidema.
\newblock Quantifying attention flow in transformers.
\newblock In Dan Jurafsky, Joyce Chai, Natalie Schluter, and Joel Tetreault, editors, {\em Proceedings of the 58th Annual Meeting of the Association for Computational Linguistics}, pages 4190--4197, Online, July 2020. Association for Computational Linguistics.

\bibitem{nsd}
Emily~J Allen, Ghislain St-Yves, Yihan Wu, Jesse~L Breedlove, Jacob~S Prince, Logan~T Dowdle, Matthias Nau, Brad Caron, Franco Pestilli, Ian Charest, et~al.
\newblock A massive 7t fmri dataset to bridge cognitive neuroscience and artificial intelligence.
\newblock {\em Nature neuroscience}, 25(1):116--126, 2022.

\bibitem{baldassarre2012individual}
Antonello Baldassarre, Christopher~M Lewis, Giorgia Committeri, Abraham~Z Snyder, Gian~Luca Romani, and Maurizio Corbetta.
\newblock Individual variability in functional connectivity predicts performance of a perceptual task.
\newblock {\em Proceedings of the National Academy of Sciences}, 109(9):3516--3521, 2012.

\bibitem{emb}
Omar Chehab, Alexandre D{\'e}fossez, Loiseau Jean-Christophe, Alexandre Gramfort, and Jean-Remi King.
\newblock Deep recurrent encoder: an end-to-end network to model magnetoencephalography at scale.
\newblock {\em Neurons, Behavior, Data Analysis, and Theory}, 2022.

\bibitem{mind-vis}
Zijiao Chen, Jiaxin Qing, Tiange Xiang, Wan~Lin Yue, and Juan~Helen Zhou.
\newblock Seeing beyond the brain: Conditional diffusion model with sparse masked modeling for vision decoding.
\newblock In {\em Proceedings of the IEEE/CVF Conference on Computer Vision and Pattern Recognition}, pages 22710--22720, 2023.

\bibitem{brain_region2}
Hans P~Op de~Beeck, Ineke Pillet, and J~Brendan Ritchie.
\newblock Factors determining where category-selective areas emerge in visual cortex.
\newblock {\em Trends in cognitive sciences}, 23(9):784--797, 2019.

\bibitem{dittadi2021transfer}
Andrea Dittadi, Frederik Tr{\"a}uble, Francesco Locatello, Manuel W{\"u}thrich, Vaibhav Agrawal, Ole Winther, Stefan Bauer, and Bernhard Sch{\"o}lkopf.
\newblock On the transfer of disentangled representations in realistic settings.
\newblock In {\em 9th International Conference on Learning Representations}, 2021.

\bibitem{dubois2016building}
Julien Dubois and Ralph Adolphs.
\newblock Building a science of individual differences from fmri.
\newblock {\em Trends in cognitive sciences}, 20(6):425--443, 2016.

\bibitem{brain_region1}
Scott~L Fairhall and Alfonso Caramazza.
\newblock Brain regions that represent amodal conceptual knowledge.
\newblock {\em Journal of Neuroscience}, 33(25):10552--10558, 2013.

\bibitem{freesurfer}
Bruce Fischl.
\newblock Freesurfer.
\newblock {\em Neuroimage}, 62(2):774--781, 2012.

\bibitem{fsaverage}
Bruce Fischl, Martin~I Sereno, Roger~BH Tootell, and Anders~M Dale.
\newblock High-resolution intersubject averaging and a coordinate system for the cortical surface.
\newblock {\em Human brain mapping}, 8(4):272--284, 1999.

\bibitem{fumero2023leveraging}
Marco Fumero, Florian Wenzel, Luca Zancato, Alessandro Achille, Emanuele Rodol{\`a}, Stefano Soatto, Bernhard Sch{\"o}lkopf, and Francesco Locatello.
\newblock Leveraging sparse and shared feature activations for disentangled representation learning.
\newblock {\em Advances in Neural Information Processing Systems}, 36:27682--27698, 2023.

\bibitem{brain_people1}
Michelle~R Greene and Aude Oliva.
\newblock The briefest of glances: The time course of natural scene understanding.
\newblock {\em Psychological science}, 20(4):464--472, 2009.

\bibitem{grill2004human}
Kalanit Grill-Spector and Rafael Malach.
\newblock The human visual cortex.
\newblock {\em Annu. Rev. Neurosci.}, 27(1):649--677, 2004.

\bibitem{data_prep1}
Zijin Gu, Keith Jamison, Amy Kuceyeski, and Mert~R. Sabuncu.
\newblock Decoding natural image stimuli from fmri data with a surface-based convolutional network.
\newblock In Ipek Oguz, Jack Noble, Xiaoxiao Li, Martin Styner, Christian Baumgartner, Mirabela Rusu, Tobias Heinmann, Despina Kontos, Bennett Landman, and Benoit Dawant, editors, {\em Medical Imaging with Deep Learning}, volume 227 of {\em Proceedings of Machine Learning Research}, pages 107--118. PMLR, 10--12 Jul 2024.

\bibitem{i2fmri3}
Zijin Gu, Keith~Wakefield Jamison, Meenakshi Khosla, Emily~J Allen, Yihan Wu, Ghislain St-Yves, Thomas Naselaris, Kendrick Kay, Mert~R Sabuncu, and Amy Kuceyeski.
\newblock Neurogen: activation optimized image synthesis for discovery neuroscience.
\newblock {\em NeuroImage}, 247:118812, 2022.

\bibitem{cv5}
Tanmay Gupta and Aniruddha Kembhavi.
\newblock Visual programming: Compositional visual reasoning without training.
\newblock In {\em Proceedings of the IEEE/CVF Conference on Computer Vision and Pattern Recognition}, pages 14953--14962, 2023.

\bibitem{auto-drive1}
Yiduo Hao, Sohrab Madani, Junfeng Guan, Mohammed Alloulah, Saurabh Gupta, and Haitham Hassanieh.
\newblock Bootstrapping autonomous driving radars with self-supervised learning.
\newblock In {\em Proceedings of the IEEE/CVF Conference on Computer Vision and Pattern Recognition}, pages 15012--15023, 2024.

\bibitem{brain_semantic_selectivity2}
Klaus Hoenig, Eun-Jin Sim, Viktor Bochev, B{\"a}rbel Herrnberger, and Markus Kiefer.
\newblock Conceptual flexibility in the human brain: dynamic recruitment of semantic maps from visual, motor, and motion-related areas.
\newblock {\em Journal of Cognitive Neuroscience}, 20(10):1799--1814, 2008.

\bibitem{cv6}
Yihan Hu, Jiazhi Yang, Li~Chen, Keyu Li, Chonghao Sima, Xizhou Zhu, Siqi Chai, Senyao Du, Tianwei Lin, Wenhai Wang, et~al.
\newblock Planning-oriented autonomous driving.
\newblock In {\em Proceedings of the IEEE/CVF Conference on Computer Vision and Pattern Recognition}, pages 17853--17862, 2023.

\bibitem{brain_semantic_selectivity1}
Alexander~G Huth, Shinji Nishimoto, An~T Vu, and Jack~L Gallant.
\newblock A continuous semantic space describes the representation of thousands of object and action categories across the human brain.
\newblock {\em Neuron}, 76(6):1210--1224, 2012.

\bibitem{disease2}
Kyungdo Kim, Sihan Lyu, Sneha Mantri, and Timothy~W Dunn.
\newblock Tulip: Multi-camera 3d precision assessment of parkinson's disease.
\newblock In {\em Proceedings of the IEEE/CVF Conference on Computer Vision and Pattern Recognition}, pages 22551--22562, 2024.

\bibitem{crowdsourcing}
Adriana Kovashka, Olga Russakovsky, Li~Fei-Fei, Kristen Grauman, et~al.
\newblock Crowdsourcing in computer vision.
\newblock {\em Foundations and Trends{\textregistered} in computer graphics and Vision}, 10(3):177--243, 2016.

\bibitem{cv4}
Zhengqi Li, Richard Tucker, Noah Snavely, and Aleksander Holynski.
\newblock Generative image dynamics.
\newblock In {\em Proceedings of the IEEE/CVF Conference on Computer Vision and Pattern Recognition}, pages 24142--24153, 2024.

\bibitem{cv3}
Youwei Liang, Junfeng He, Gang Li, Peizhao Li, Arseniy Klimovskiy, Nicholas Carolan, Jiao Sun, Jordi Pont-Tuset, Sarah Young, Feng Yang, et~al.
\newblock Rich human feedback for text-to-image generation.
\newblock In {\em Proceedings of the IEEE/CVF Conference on Computer Vision and Pattern Recognition}, pages 19401--19411, 2024.

\bibitem{mindreader}
Sikun Lin, Thomas Sprague, and Ambuj~K Singh.
\newblock Mind reader: Reconstructing complex images from brain activities.
\newblock {\em Advances in Neural Information Processing Systems}, 35:29624--29636, 2022.

\bibitem{mscoco}
Tsung-Yi Lin, Michael Maire, Serge Belongie, James Hays, Pietro Perona, Deva Ramanan, Piotr Doll{\'a}r, and C~Lawrence Zitnick.
\newblock Microsoft coco: Common objects in context.
\newblock In {\em Computer Vision--ECCV 2014: 13th European Conference, Zurich, Switzerland, September 6-12, 2014, Proceedings, Part V 13}, pages 740--755. Springer, 2014.

\bibitem{brain-dive}
Andrew Luo, Maggie Henderson, Leila Wehbe, and Michael Tarr.
\newblock Brain diffusion for visual exploration: Cortical discovery using large scale generative models.
\newblock {\em Advances in Neural Information Processing Systems}, 36, 2024.

\bibitem{brain-scuba}
Andrew~F Luo, Margaret~M Henderson, Michael~J Tarr, and Leila Wehbe.
\newblock Brainscuba: Fine-grained natural language captions of visual cortex selectivity.
\newblock {\em arXiv preprint arXiv:2310.04420}, 2023.

\bibitem{mangun1998erp}
George~R Mangun, Michael~H Buonocore, Massimo Girelli, and Amishi~P Jha.
\newblock Erp and fmri measures of visual spatial selective attention.
\newblock {\em Human brain mapping}, 6(5-6):383--389, 1998.

\bibitem{mcc}
Brian~W Matthews.
\newblock Comparison of the predicted and observed secondary structure of t4 phage lysozyme.
\newblock {\em Biochimica et Biophysica Acta (BBA)-Protein Structure}, 405(2):442--451, 1975.

\bibitem{brain_plasticity}
Arne May.
\newblock Experience-dependent structural plasticity in the adult human brain.
\newblock {\em Trends in cognitive sciences}, 15(10):475--482, 2011.

\bibitem{mni152}
John Mazziotta, Arthur Toga, Alan Evans, Peter Fox, Jack Lancaster, Karl Zilles, Roger Woods, Tomas Paus, Gregory Simpson, Bruce Pike, et~al.
\newblock A four-dimensional probabilistic atlas of the human brain.
\newblock {\em Journal of the American Medical Informatics Association}, 8(5):401--430, 2001.

\bibitem{phan2002functional}
K~Luan Phan, Tor Wager, Stephan~F Taylor, and Israel Liberzon.
\newblock Functional neuroanatomy of emotion: a meta-analysis of emotion activation studies in pet and fmri.
\newblock {\em Neuroimage}, 16(2):331--348, 2002.

\bibitem{disease1}
Vu~Minh~Hieu Phan, Yutong Xie, Yuankai Qi, Lingqiao Liu, Liyang Liu, Bowen Zhang, Zhibin Liao, Qi~Wu, Minh-Son To, and Johan~W Verjans.
\newblock Decomposing disease descriptions for enhanced pathology detection: A multi-aspect vision-language pre-training framework.
\newblock In {\em Proceedings of the IEEE/CVF Conference on Computer Vision and Pattern Recognition}, pages 11492--11501, 2024.

\bibitem{clip}
Alec Radford, Jong~Wook Kim, Chris Hallacy, Aditya Ramesh, Gabriel Goh, Sandhini Agarwal, Girish Sastry, Amanda Askell, Pamela Mishkin, Jack Clark, et~al.
\newblock Learning transferable visual models from natural language supervision.
\newblock In {\em International conference on machine learning}, pages 8748--8763. PMLR, 2021.

\bibitem{rec_by_gan2}
N~Apurva Ratan~Murty, Pouya Bashivan, Alex Abate, James~J DiCarlo, and Nancy Kanwisher.
\newblock Computational models of category-selective brain regions enable high-throughput tests of selectivity.
\newblock {\em Nature communications}, 12(1):5540, 2021.

\bibitem{mindeye1}
Paul Scotti, Atmadeep Banerjee, Jimmie Goode, Stepan Shabalin, Alex Nguyen, Aidan Dempster, Nathalie Verlinde, Elad Yundler, David Weisberg, Kenneth Norman, et~al.
\newblock Reconstructing the mind's eye: fmri-to-image with contrastive learning and diffusion priors.
\newblock {\em Advances in Neural Information Processing Systems}, 36, 2024.

\bibitem{mindeye2}
Paul~Steven Scotti, Mihir Tripathy, Cesar Torrico, Reese Kneeland, Tong Chen, Ashutosh Narang, Charan Santhirasegaran, Jonathan Xu, Thomas Naselaris, Kenneth~A Norman, et~al.
\newblock Mindeye2: Shared-subject models enable fmri-to-image with 1 hour of data.
\newblock In {\em Forty-first International Conference on Machine Learning}, 2024.

\bibitem{gradcam}
Ramprasaath~R Selvaraju, Michael Cogswell, Abhishek Das, Ramakrishna Vedantam, Devi Parikh, and Dhruv Batra.
\newblock Grad-cam: Visual explanations from deep networks via gradient-based localization.
\newblock In {\em Proceedings of the IEEE international conference on computer vision}, pages 618--626, 2017.

\bibitem{cv2}
Samuel Stevens, Jiaman Wu, Matthew~J Thompson, Elizabeth~G Campolongo, Chan~Hee Song, David~Edward Carlyn, Li~Dong, Wasila~M Dahdul, Charles Stewart, Tanya Berger-Wolf, et~al.
\newblock Bioclip: A vision foundation model for the tree of life.
\newblock In {\em Proceedings of the IEEE/CVF Conference on Computer Vision and Pattern Recognition}, pages 19412--19424, 2024.

\bibitem{data_prep2}
Yu~Takagi and Shinji Nishimoto.
\newblock High-resolution image reconstruction with latent diffusion models from human brain activity.
\newblock In {\em Proceedings of the IEEE/CVF Conference on Computer Vision and Pattern Recognition}, pages 14453--14463, 2023.

\bibitem{trauble2021disentangled}
Frederik Tr{\"a}uble, Elliot Creager, Niki Kilbertus, Francesco Locatello, Andrea Dittadi, Anirudh Goyal, Bernhard Sch{\"o}lkopf, and Stefan Bauer.
\newblock On disentangled representations learned from correlated data.
\newblock In {\em International conference on machine learning}, pages 10401--10412. PMLR, 2021.

\bibitem{fmri_noise1}
Kamil U{\u{g}}urbil, Junqian Xu, Edward~J Auerbach, Steen Moeller, An~T Vu, Julio~M Duarte-Carvajalino, Christophe Lenglet, Xiaoping Wu, Sebastian Schmitter, Pierre~Francois Van~de Moortele, et~al.
\newblock Pushing spatial and temporal resolution for functional and diffusion mri in the human connectome project.
\newblock {\em Neuroimage}, 80:80--104, 2013.

\bibitem{mind-bridge}
Shizun Wang, Songhua Liu, Zhenxiong Tan, and Xinchao Wang.
\newblock Mindbridge: A cross-subject brain decoding framework.
\newblock In {\em Proceedings of the IEEE/CVF Conference on Computer Vision and Pattern Recognition}, pages 11333--11342, 2024.

\bibitem{brain_people2}
Galit Yovel and Nancy Kanwisher.
\newblock The neural basis of the behavioral face-inversion effect.
\newblock {\em Current biology}, 15(24):2256--2262, 2005.

\bibitem{cv1}
Zehao Yu, Anpei Chen, Binbin Huang, Torsten Sattler, and Andreas Geiger.
\newblock Mip-splatting: Alias-free 3d gaussian splatting.
\newblock In {\em Proceedings of the IEEE/CVF Conference on Computer Vision and Pattern Recognition}, pages 19447--19456, 2024.

\bibitem{auto-drive2}
Jimuyang Zhang, Zanming Huang, Arijit Ray, and Eshed Ohn-Bar.
\newblock Feedback-guided autonomous driving.
\newblock In {\em Proceedings of the IEEE/CVF Conference on Computer Vision and Pattern Recognition}, pages 15000--15011, 2024.

\bibitem{zhang2024identifiable}
Qi~Zhang, Yifei Wang, and Yisen Wang.
\newblock Identifiable contrastive learning with automatic feature importance discovery.
\newblock {\em Advances in Neural Information Processing Systems}, 36, 2024.

\bibitem{clip_mused}
Qiongyi Zhou, Changde Du, Shengpei Wang, and Huiguang He.
\newblock Clip-mused: Clip-guided multi-subject visual neural information semantic decoding.
\newblock In {\em The Twelfth International Conference on Learning Representations}, 2024.

\bibitem{auto-drive3}
Xiaoyu Zhou, Zhiwei Lin, Xiaojun Shan, Yongtao Wang, Deqing Sun, and Ming-Hsuan Yang.
\newblock Drivinggaussian: Composite gaussian splatting for surrounding dynamic autonomous driving scenes.
\newblock In {\em Proceedings of the IEEE/CVF Conference on Computer Vision and Pattern Recognition}, pages 21634--21643, 2024.

\end{thebibliography}
